\documentclass[english,aps,prstper,reprint,showpacs,longbibliography,superscriptaddress,nofootinbib]{revtex4-2}   

\usepackage[T1]{fontenc}	
\usepackage[latin9]{inputenc}	
\usepackage{geometry}		
\usepackage{graphicx}
\usepackage{times}
\usepackage{pifont}

\usepackage{hyperref}  
\hypersetup{colorlinks=true,urlcolor=blue,citecolor=blue,linkcolor=blue}   
\usepackage[caption=false]{subfig}
\usepackage[shortlabels]{enumitem}
\setlist{nosep} 

\begin{document}

\begin{titlepage}

  
  \title{Comparing intro and beyond-intro students' reasoning about uncertainty}

  \author{Emily M. Stump}
  \affiliation{Laboratory of Atomic and Solid State Physics, Cornell University, 245 East Avenue, Ithaca, NY,
    14853} 
  \author{Mark Hughes}
  \affiliation{Department of Physics, California State University Fullerton, 800 N. State College Blvd., Fullerton, CA,
    92831} 
  \author{Gina Passante}
  \affiliation{Department of Physics, California State University Fullerton, 800 N. State College Blvd., Fullerton, CA,
    92831} 
   \author{N. G. Holmes}
   \affiliation{Laboratory of Atomic and Solid State Physics, Cornell University, 245 East Avenue, Ithaca, NY,
    14853} 


  \begin{abstract}
    Uncertainty is an important concept in physics laboratory instruction. However, little work has examined how students reason about uncertainty beyond the introductory (intro) level. In this work we aimed to compare intro and beyond-intro students' ideas about uncertainty. We administered a survey to students at 10 different universities with questions probing procedural reasoning about measurement, student-identified sources of uncertainty, and predictive reasoning about data distributions. We found that intro and beyond-intro students answered similarly on questions where intro students already exhibited expert-level reasoning, such as in comparing two data sets with the same mean but different spreads, identifying limitations in an experimental setup, and predicting how a data distribution would change if more data were collected. For other questions, beyond-intro students generally exhibited more expert-like reasoning than intro students, such as when determining whether two sets of data agree, identifying principles of measurement that contribute to spread, and predicting how a data distribution would change if better data were collected. Neither differences in student populations, lab courses taken, nor research experience were able to fully explain the variability between intro and beyond-intro student responses. These results call for further research to better understand how students' ideas about uncertainty develop beyond the intro level.
  \end{abstract}
  
  \maketitle
\end{titlepage}

\section{Introduction}

Laboratory instruction is a key part of the undergraduate physics curriculum, providing students with the opportunity to develop experimental skills and knowledge not covered in theory-focused courses~\cite{Kozminski2014}. One of these experimental skills is the ability to make decisions about interpreting data and drawing conclusions from an experiment~\cite{Walsh2019,Holmes2019,Holmes2015a,Holmes2015}. Integral to developing this decision-making skill is understanding measurement uncertainty~\cite{Allie2003,Kung2005,Buffler2008}. While many studies have probed introductory students' ideas about uncertainty, very little work has probed students' understanding of uncertainty beyond the intro level. This paper aims to bridge this gap by probing both intro and beyond-intro students' reasoning about different aspects of uncertainty.

\subsection{Student understanding of uncertainty at the intro level}

Most of the research on student understanding of uncertainty has focused on procedural reasoning: given some data, what (if any) additional data should they collect and/or how should they analyze the data they have. In a study of first-year undergraduate students, S\'{e}r\'{e} \textit{et al.}~\cite{Sere1993} found that most students struggled to use multiple measurements in interpreting their data. Most students would take only a single measurement in a lab experiment and would take more measurements only if explicitly prompted to do so. If required to take more measurements, students trusted the first measurement more than the subsequent ones and classified each measurement as ``good'' or ``bad,'' rather than seeing the entire set of measurements as valuable. In a similar study, Coelho and S\'{e}r\'{e}~\cite{Coelho1998} found that high school students also emphasized individual measurements, believing that an experiment has a ``true value'' associated with it that they should be able to determine with a single perfect measurement. Because of this emphasis on an idealized single measurement, students tended not to consider uncertainty when evaluating whether two measurements were identical~\cite{Sere1993,Coelho1998}. Other studies of intro undergraduate students found similar reasoning across science disciplines~\cite{Leach1998,Evangelinos1999}. A significant portion of students believed that it is possible to make a perfect measurement of a ``true value'' with sufficient time and money. Many students also focused on the average value in comparing two data sets and did not consider the spread in the data.

These findings were replicated and extended by Allie, Buffler, Campbell, Lubben, and colleagues studying intro-level students at the University of Cape Town in South Africa~\cite{Allie1998,Buffler2001,Lubben2001}. These researchers developed a survey instrument to probe students' procedural reasoning related to measurement and uncertainty: the Physics Measurement Questionnaire (PMQ). Based on the student responses, they identified two main paradigms of student procedural reasoning: point and set reasoning. Point reasoners tend to emphasize individual measurements in interpreting data. They believe that any single measurement could yield the ``true value'' of a parameter and that deviation from the ``true value'' results from mistakes in the experiment. These students see taking multiple measurements as beneficial solely as a way to practice so that they can ultimately make a single perfect measurement. In contrast, set reasoners view each individual measurement as an estimate of the quantity of interest. They regard uncertainty as a natural part of experimentation and consequently rely on a \textit{set} of measurements when interpreting their data. They use statistics such as the mean and standard deviation when reporting on their data, as opposed to reporting an individual measured value. The set paradigm is seen as the ``expertlike'' approach to measurement and uncertainty, and, accordingly, the goal of instruction is argued to be to shift students' reasoning away from the point paradigm and toward the set paradigm~\cite{Lubben2001,Buffler2001}.

These researchers used the PMQ and the point and set paradigms to probe intro-level students' reasoning, both before~\cite{Lubben2001} and after~\cite{Buffler2001} taking a lab course. Lubben \textit{et al.}~\cite{Lubben2001} found that prior to instruction, the students' responses were split approximately equally between point and set reasoning for questions about making more than one measurement, with point reasoners arguing a single data point was sufficient and set reasoners arguing in favor of collecting more data. In comparing two data sets, however, nearly all students relied on comparing the average values (point) rather than taking into account the spread in the data (set). After instruction, the majority of students exhibited set reasoning in the questions about taking more data, but most still struggled to consistently apply set reasoning in comparing data sets~\cite{Buffler2001}. Like the students in S\'{e}r\'{e} \textit{et al.}'s~\cite{Sere1993} study, these students struggled to understand how uncertainty and spread in data should be used when interpreting experimental results.

Since then, additional researchers have measured the efficacy of various traditional lab courses at teaching set reasoning or have developed intro lab courses with the goal of shifting students away from point reasoning toward set reasoning, with mixed success. Although students have exhibited increases in set reasoning on the PMQ and similar questions from pre-instruction to post-instruction, many (or, in some cases, most) students still apply ``mixed'' reasoning, using set reasoning in some contexts but point reasoning in others, particularly in comparing two sets of data~\cite{Kung2005,Volkwyn2008,Wan2022,Pollard2020,Kung2006,Pillay2008}. These results indicate that a single intro-level lab course is likely insufficient for undergraduate students to master procedural reasoning about uncertainty.

Several studies have also probed students' ideas about sources of uncertainty and how these relate to students' procedural reasoning. For example, previous studies have found that while students are able to identify various sources of uncertainty in an experiment, they may struggle to appropriately quantify those sources~\cite{Etkina2008,Sere1993}. Moreover, researchers have expressed concern about students' tendency to attribute uncertainty exclusively to something going wrong in the experiment or ``human error''~\cite{Hu2018,Holmes2015,Sere2001,Evangelinos2002,Allie2003}. The concern centers on observations that this conception of uncertainty can lead students to believe that all uncertainty can be eliminated in an experiment, which is aligned with point reasoning. To address this issue, researchers have advocated for teaching uncertainty not as a list of mistakes to be fixed but rather as a fundamental aspect of experimentation that must be quantified and used to interpret data~\cite{Evangelinos1999,Evangelinos2002,Allie2003,Pillay2008,Buffler2008}.

\subsection{Student understanding of uncertainty beyond the intro level}

To our knowledge, only a handful of studies have investigated students' ideas about uncertainty beyond the intro level. Hu and Zwickl~\cite{Hu2018} probed intro-level, upper-level, and Ph.D. students' views about uncertainty. They observed that upper-level students were more likely than intro-level students to identify uncertainty evaluation as important for evaluating whether an experimental result is trustworthy. They also found that Ph.D. students and upper-level students were more likely than intro-level students to view the purpose of uncertainty analysis as quantifying reliability, although only Ph.D. students flagged uncertainty as an inherent aspect of experimentation. Overall, these results suggest that students' views of measurement can change greatly throughout the physics curriculum.

Our previous work~\cite{Stein2019,Stump2020,White2020,StumpTBD,SchangTBD} has focused on upper-level students' ideas about sources of uncertainty and predictive reasoning about uncertainty in different experimental contexts. We found that students were more likely to identify physics principles as sources of uncertainty in quantum-mechanics experiments (e.g., the single-slit experiment) than in classical experiments (e.g., projectile motion) and in experiments with a theoretical expected distribution of outcomes (e.g., Brownian motion) than in experiments with a theoretical single outcome (e.g., projectile motion)~\cite{Stein2019,Stump2020,StumpTBD}. We also asked students about how a data distribution would change if an experiment were performed by a larger group of students (more data) or if an experiment were performed by experts (better data)~\cite{Stein2019,White2020,SchangTBD}. Most students indicated that more data would result in the same distribution (the ``correct answer''), although a sizeable minority indicated that more data would result in a narrower distribution. For better data, students indicated that experts would either measure the same distribution or measure a narrower distribution, with students more likely to answer ``same'' for quantum-mechanics experiments (e.g. the single-slit experiment) and in experiments with a theoretical expected distribution of outcomes (e.g. Brownian motion).

\subsection{Research aims}

Given that many students leave intro-level lab courses with at least some point-like ideas about uncertainty~\cite{Kung2005,Volkwyn2008,Wan2022,Pollard2020,Kung2006,Pillay2008} and the dearth of research on student ideas about uncertainty beyond the intro level, our goal was to broadly characterize the reasoning of a diverse sample of intro and beyond-intro students using previously developed measures of student thinking. In particular, we probed three aspects of student reasoning: procedural reasoning~\cite{Allie1998,Buffler2001,Lubben2001}, ideas about sources of uncertainty~\cite{StumpTBD}, and predictive reasoning about measuring more or better data~\cite{SchangTBD}. We found that intro students were already expertlike in their reasoning on two of the five questions, but that beyond-intro students were more expertlike than intro students on the other three probes. We tested (and ruled out) several plausible explanations for the observed differences (selection based on major, lab course experience, research experience, and institutional variability). We use prior work to further situate these results and propose future research directions.

\section{Methods}

In this section we describe the survey questions we analyzed and the data collection process. We also describe the coding schemes used to interpret open-ended questions and our approach to making quantitative claims.

\subsection{Survey questions}

The survey used in this work is adapted from surveys used in previous work probing student reasoning about uncertainty~\cite{SchangTBD,StumpTBD,Allie1998}. The survey questions analyzed here center around a single experimental scenario of a ball rolling down a ramp that was adapted from the Physics Measurement Questionnaire (PMQ)~\cite{Allie1998}. Although student reasoning about measurement can vary significantly across different experimental contexts or in generalized questions~\cite{Leach2000, StumpTBD, SchangTBD}, this scenario provides a snapshot of intro and beyond-intro students' reasoning in the context of a familiar experiment and allows us to compare our results to previous findings from the PMQ.

Students are first asked two questions from the PMQ: the same mean, different spread probe (SMDS) and the different mean, same spread probe (DMSS) (see Figs.~\ref{fig:scenario},~\ref{fig:smds}, and~\ref{fig:dmss} in the appendix). In the SMDS probe, two groups of students have each measured five data points such that the two groups have the same mean value but different spread in their data. The probe presents three possible viewpoints about which group has achieved better results and survey respondents are asked to identify with which viewpoint they agree and explain why. The DMSS probe is set up similarly: two groups have each measured five data points, but this time the groups have different mean values. The probe presents two possible viewpoints on whether the two groups' results agree and survey respondents are again asked to identify with which viewpoint they agree and why.

\begin{figure*}
    \centering
    \includegraphics[width=.9\textwidth]{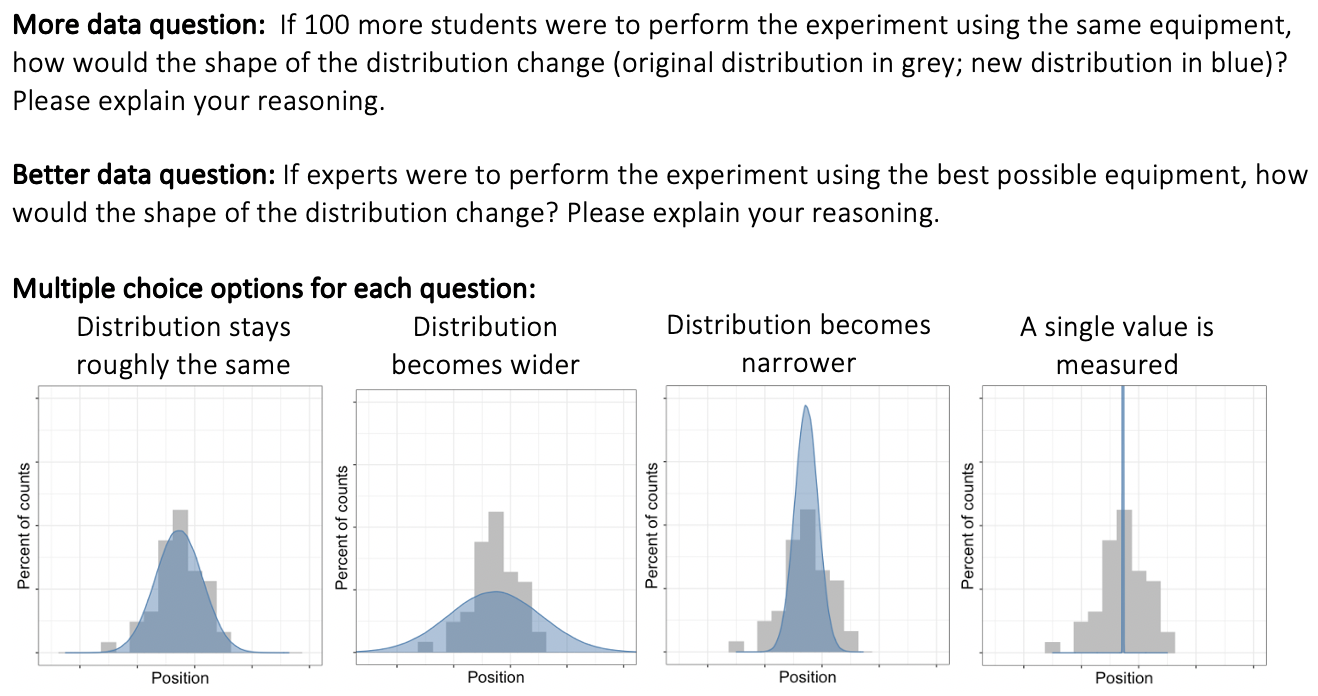}
    \caption{Text and multiple-choice options for the More Data and Better Data questions~\cite{SchangTBD}.}
    \label{fig:more_better_questions}
\end{figure*}

The survey then presents a fictitious histogram of measurements collected by 50 students in a lab class (see Ref.~\cite{StumpTBD}). Students are asked to list sources of uncertainty that contribute to the spread in the data (Sources):
\begin{quote}
    ``What is causing the shape of the distribution? List as many causes as you can think of.''
\end{quote}
Finally, students are asked two closed-response questions about how the fictitious histogram would change if either 100 more students (More Data) or experts using the best possible equipment (Better Data) performed the experiment~\cite{SchangTBD} (see Fig.~\ref{fig:more_better_questions}).

The questions were always asked in the same order: SMDS, DMSS, Sources, More Data, Better Data. Students could return to previous questions to change their answers as they progressed through the survey.

At the end of the survey, students were also asked a series of questions related to demographics (race/ethnicity, gender, major, etc.) and educational experiences. We asked students to report which types of lab courses they had taken:
\begin{quote}
    ``Are you currently taking or have you previously taken any of the following types of college physics lab class? Choose all that apply.''
    \begin{itemize}
        \item[\ding{114}] Introductory mechanics and/or E\&M and/or waves/thermo
        \item[\ding{114}] Upper division: electronics
        \item[\ding{114}] Upper division: optics
        \item[\ding{114}] Advanced lab
        \item[\ding{114}] Other: \_\_\_\_\_\_\_\_
    \end{itemize}
\end{quote}
We also asked students to report whether they had research experience:
\begin{quote}
Are you currently conducting or have you previously conducted research in any of the following areas? Choose all that apply.
\begin{itemize}
    \item[\ding{114}] Experimental physics or astrophysics
    \item[\ding{114}] Theoretical physics or astrophysics
    \item[\ding{114}] Computational physics or astrophysics
    \item[\ding{114}] Experimental research in another science
    \item[\ding{114}] Theoretical research in another science
    \item[\ding{114}] Computational research in another science
    \item[\ding{114}] Other: \_\_\_\_\_\_\_\_
    \item[\ding{114}] No research experience
\end{itemize}
\end{quote}

We use these questions to report on how many students had taken at least one lab course beyond the intro level (marking at least one of Upper division: electronics, Upper division: optics, Advanced lab, or Other (if applicable)) and how many students had experimental research experience (marking at least one of Experimental physics or astrophysics or Experimental research in another science).

\subsection{Data collection}

The survey was administered online via Qualtrics during the second half of the Fall 2021 and Spring 2022 semesters. We targeted students who were either enrolled in introductory mechanics or electricity and magnetism courses (``intro'' students) or who were enrolled in or had taken at least one quantum mechanics course (``beyond-intro'' students). In total we received survey responses from students at 10 universities, with 427 completed responses from intro students and 158 completed responses from beyond-intro students. The universities include private universities, public universities, primarily white institutions, Hispanic-serving institutions, and a historically Black university. Students were recruited by their course instructors and were offered either course credit or a drawing for a \$25 gift card for completing the survey. The numbers of participants from each university at each level are shown in Table~\ref{ta:universities} and the students' self-reported demographic information is shown in Table~\ref{ta:demographics} in the appendix. 
\begin{table} 
  \caption{Number of responses from intro ($N=427$) and beyond-intro ($N=158$) students by university.}
  \label{ta:universities}
  \begin{ruledtabular}
    \begin{tabular}{lrr}
    Institution & Intro & Beyond-intro\\
    \hline
    Auburn University & 99 & 0\\
    California State Polytechnic University &&\\
    \quad Pomona & 3 & 0\\
    California State University Fullerton & 3 & 0\\
    California State University San Marcos & 26 & 19\\
    Cornell University & 119 & 43\\
    North Carolina A\&T State University & 89 & 0\\
    San Jos\'{e} State University & 0 & 10\\
    Texas A\&M University & 88 & 7\\
    University of Colorado Boulder & 0 & 78\\
    University of Wisconsin Stout & 0 & 1
    \end{tabular}
  \end{ruledtabular}
\end{table}

We also asked students questions related to their majors, lab courses taken, and research experience. The intro students were primarily non-physics majors, with 59\% engineering majors and 14\% other STEM majors; only 15\% were physics majors. These students were either enrolled in an intro lab course (49\%) or had not yet taken a physics lab course in college. Only 9\% of intro students had experimental research experience. The beyond-intro students, in contrast, were primarily physics majors (92\%). These students had more experience with lab courses and experimental physics compared to the intro students, with 56\% having taken at least one lab course beyond the intro level and 41\% having experimental research experience.

\subsection{Coding schemes}

The SMDS, DMSS, and sources of uncertainty questions were all open-ended. Thus, we used established coding schemes to characterize student responses to these questions.

The coding schemes for the SMDS and DMSS questions were developed based on the coding schemes in the original papers about the PMQ~\cite{Allie1998,Buffler2001,Lubben2001} and modified slightly to reflect the student responses in our sample. Responses were coded based on their alignment with the point and set paradigms. The point code was given to responses that focused on comparing only the average values for the data sets or that compared the individual data points in the two data sets rather than considering the spread in the data. For example, point responses may argue that the spread is irrelevant to the quality of data or use the percent error between the means to decide whether two data sets agree. The set code was given to responses that mentioned the spread in the data as the justification for the selected viewpoint. For example, set responses may argue that the group with the smaller spread in their data had a better result in the SMDS probe or that the two means in DMSS probe agreed because there was significant overlap in the spreads of the two data sets. Responses that could not be coded as either point or set or that contained elements of both point and set reasoning were coded as unclear (as in Refs.~\cite{Allie1998, Buffler2001, Lubben2001}). Example point and set responses are shown in Table~\ref{ta:pmq}.

\begin{table*}[tb] 
  \caption{Examples of responses receiving the point and set codes from the SMDS and DMSS probes.}
  \label{ta:pmq}
  \begin{ruledtabular}
    \begin{tabular}{p{0.03\textwidth}p{0.46\textwidth}p{0.46\textwidth}}
    \textbf{Code} & \textbf{SMDS responses}& \textbf{DMSS responses}\\
      \hline
    Point & \textit{``As long as the average is the same, then the right procedures were followed''} (intro student) & \textit{``Even though they are close, the averages are not the same.''} (intro student)\\
    & \textit{``I agree most with B because the point of an experiment is not to get the result that is better, but the one that is closely related to the values that the students are predicting. Even though the distribution of their data are different, they still have the same average so I would say that both are similar.''} (intro student) & \textit{``They are extremely similar values, just a few different points were varied. The only difference in the ending values came from slight differences in measurements throughout the experiment, which is normal due to human error.''} (intro student)\\
    & \textit{``By only looking at the data, I don't believe group B has better results than group A or vice versa. I say this because they both have the same average of 435 mm.''} (beyond-intro student) & \textit{``The two averages have a percentage difference of 0.46\%, so they agree.''} (beyond-intro student)\\
    & \textit{``I don't really understand what any of these groups mean by 'better'. These are just data from an experiment and just because the data has a greater standard deviation (or spread) it does not mean that it is any better or worse than another set of data.''} (beyond-intro student) & \textit{``Although the average is not the same, they are very close, averages improve over time and do not have to be equal for agreement''} (beyond-intro student)\\
    \hline
    Set & \textit{``The data for A is less spread because when looking a the table, the highest and lowest values have a smaller difference compared to B's. A's difference is only 20 mm and B's is 50.''} (intro student) & \textit{``The distribution of each group's data is similar enough to confidently assume that the data is reflecting the same phenomena.''} (intro student) \\
    &\textit{``Although the average is the same, the lower range of the first group is superior.''} (intro student) &  \textit{``Both groups should construct confidence intervals. They appear as though they will overlap and therefore agree with each other.''} (intro student)\\
    & \textit{``I agree with A, because the numbers when measured are closer together, which means their measurements are more precise.''} (beyond-intro student) & \textit{``The difference in their means is within the range allowable by the variance of their data''} (beyond-intro student)\\
    & \textit{``Group A had more consistent data points with a lower deviation. This indicates that the experiment was done more consistently and carefully, resulting in a lower error than group B.
Because group A has a lower estimated error than group B, group A can be more confident in their end average than group B, if each group uses only their own data.''} (beyond-intro student)& \textit{``The standard error of the mean of each group is about the same, about 6 mm. The averages are within one standard deviation of one another, so the results agree.''} (beyond-intro student) \\
    \end{tabular}
  \end{ruledtabular}
\end{table*}

Two of the authors independently coded a random sample of 50 responses from the SMDS probe and 50 responses from the DMSS probe. We quantified inter-rater reliability using Cohen's kappa, achieving values of 0.92 and 0.8 for the SMDS and DMSS probes, respectively, which indicates almost perfect agreement~\cite{Landis1977}. The two researchers then split the remaining responses and independently coded them.

Responses to the Sources question were coded using a previously developed coding scheme~\cite{StumpTBD}. This coding scheme classifies student-listed sources of uncertainty as limitations or principles and is based on the Modeling Framework for Experimental Physics~\cite{Zwickl2015,DounasFrazer2018}. The limitations code was applied to sources of uncertainty related to imperfections in the experimental procedure or setup, such as human error in conducting an experiment, environmental factors such as air resistance, or the precision limit of a measurement device.

The principles code encapsulates both statistical principles and theoretical physics principles. The first includes the idea that experimental measurement is fundamentally probabilistic and, therefore, uncertainty must be modeled using statistical principles. The second includes sources of uncertainty that are due to principles of theoretical physics, such as the Heisenberg uncertainty principle, which places limits on the precision of measurement due to the nature of quantum-mechanical systems. In the context of this study, virtually all responses that received the principles code were related to modeling uncertainty using statistical principles. Examples of student-listed sources of uncertainty that were coded as limitations and principles are shown in Table~\ref{ta:sources}.

\begin{table*}[tb] 
  \caption{Examples of responses receiving the limitations and principles codes.}
  \label{ta:sources}
  \begin{ruledtabular}
    \begin{tabular}{p{0.05\textwidth}p{0.88\textwidth}}
    \textbf{Code} & \textbf{Examples}\\
      \hline
    Limitations & \textit{``Equipment accuracy/precision (depending on what is used to take measurements)''} (intro student)\\
    & \textit{``Different masses of balls''} (intro student) \\
    & \textit{``Slight shifting of the measuring paper before or after marking''} (beyond-intro student) \\
    & \textit{``Air currents in the room.''} (beyond-intro student) \\
    \hline
    Principles & \textit{``The distribution is roughly symmetric, and the sample size of 50 is relatively large, greater than 30. Central Limit Theorem explains that this distribution for d is approximately normal, explaining the symmetry.''} (intro student)  \\
    &\textit{``Error is typically normally distributed''} (intro student) \\
    & \textit{``in any real system, there are bound to be differences in measurements creating a normal distribution''} (beyond-intro student) \\
    & \textit{``the results are random which result in a gaussian distribution (what we observe)''} (beyond-intro student) \\
    \end{tabular}
  \end{ruledtabular}
\end{table*}

Any sources that were too vague to classify as limitations or principles were coded as unclear. Most students listed at least one source of uncertainty that we were able to code as either limitations or principles (85\% of intro students and 92\% of beyond-intro students).

The Sources coding scheme was validated in previous work~\cite{StumpTBD} in which two researchers achieved a Cohen's kappa value of 0.85, indicating almost perfect agreement~\cite{Landis1977}. One of these researchers coded all of the responses in this study.

As with any analysis of open-response surveys, both of our coding schemes are limited in scope. The codes we used are fairly broad and may fail to capture some interesting nuance in students' responses. For example, the point/set coding scheme treats each student's response as aligned either with novicelike (point) or expertlike (set) reasoning. However, there may be a range of sophistication and expertlike thinking within responses classified as point, set, or unclear that our coding scheme fails to capture. Similarly, our limitations code includes a wide variety of sources, from actionable or quantifiable sources, such as varying force applied while dropping a ball or the instrumental precision of a ruler, to the more vague and unproductive ``human error''~\cite{Hu2018,Holmes2015,Sere2001,Evangelinos2002,Allie2003}. Within this work, we were comparing student responses across five survey questions, leading to a large number of comparisons that increases the chances of seeing an effect due to random chance. As a result, we chose not to subdivide these codes further within this paper. We leave it to future work to address these nuances within our codes to further categorize student reasoning about uncertainty.

\subsection{Data analysis}

Our goal in this work was to compare the reasoning exhibited by intro and beyond-intro students and to provide possible explanations for any differences observed. For the SMDS and DMSS probes, we evaluated the fraction of intro and beyond-intro students whose response was given a code of point, set, or unclear. For the Sources question, students could list multiple sources of uncertainty, each of which received a code of limitations, principles, or unclear. We therefore compared the fractions of intro and beyond-intro students who listed at least one source that we coded as limitations and, separately, the fraction of students who listed at least one source that we coded as principles. For the more and better data probes, we evaluated the fraction of intro and beyond-intro students who chose each of the possible predicted distributions (the same, wider, narrower, or single-value).

To make quantitative comparisons between groups of students, we relied on graphical representation of the 95\% confidence interval for each proportion (estimated using the Wilson score interval~\cite{Wilson1927}). We evaluated the distinguishability of pairs of proportions based on the relative overlap of the 95\% confidence intervals. We did not quantify $p$-values due to the large number of comparisons being made and the various concerns about $p$-values in the literature~\cite[e.g.][]{cohen_earth_1994, cumming_new_2013, nosek_preregistration_2018}. 

Although there are various ways in which students may productively answer some of our survey questions, others have a clear alignment with expertlike reasoning. For the SMDS and DMSS questions, responses that are given the point code are considered to be novicelike, while responses coded as set are considered to align with expertlike reasoning~\cite{Allie1998,Buffler2001,Lubben2001}. For the Sources question, we expect that identifying principles sources of uncertainty related to the probabilistic nature of measurement is aligned with an expertlike view of measurement~\cite{Evangelinos1999,Evangelinos2002,Allie2003,Pillay2008,Buffler2008}.\footnote{Given that very few students mentioned sources related to theoretical physics principles, we do not interpret these responses.} Similarly, we consider an answer of Single value for either the More Data or Better Data question to be novicelike, as it aligns with point reasoning and is in opposition to a probabilistic understanding of measurement~\cite{Allie1998,Buffler2001,Lubben2001}. Finally, the most correct answer to the More Data question is Same, as additional students employing similar methods should measure the same distribution of results as the original students. A common incorrect answer is Narrower, as students may use a ``more data is better'' heuristic, assuming that because collecting a large amount of data is important for reducing uncertainty in the estimate of a parameter that the distribution of measurements will also become narrower~\cite{SchangTBD}.

\section{Results}

In this section we first present comparisons between intro and beyond-intro students' responses to our three types of measurement uncertainty probes. We then explore possible explanations for the differences we observed between the two populations.

\subsection{Comparing intro and beyond-intro students' reasoning}
\label{sec:overall_results}

We probed students' procedural reasoning about measurement, ideas about sources of uncertainty, and predictive reasoning about data distributions. In this section, we report on similarities and differences in reasoning between these two groups of students.

\subsubsection{Procedural reasoning}

\begin{figure}
    \centering
    \includegraphics[width=.48\textwidth]{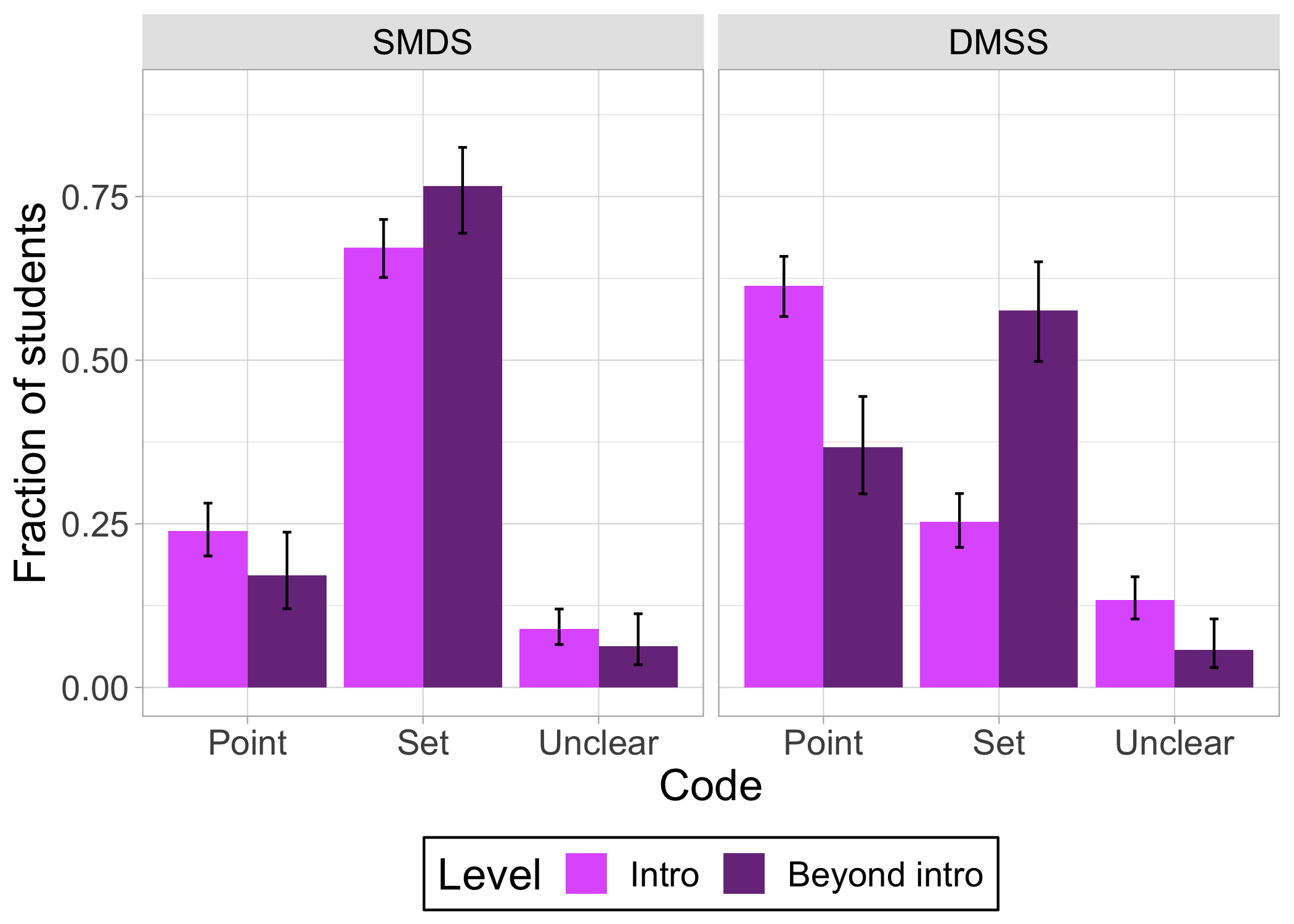}
    \caption{Point and set codes applied to intro and beyond-intro students' responses to the SMDS and DMSS probes from the PMQ~\cite{Allie1998}. Uncertainty bars represent the 95\% confidence interval.}
    \label{fig:pmq}
\end{figure}

We first asked students two questions from the PMQ~\cite{Allie1998}: the SMDS probe and the DMSS probe. The SMDS probe asks respondents to evaluate which of two data distributions with the same mean but different spreads is the better result, while the DMSS probe asks respondents to decide whether two data distributions with different means but the same spread agree. Student responses to these probes were coded as exhibiting either \textit{point} or \textit{set} reasoning; here we report on the fraction of intro and beyond-intro students' responses that received each code (see Fig.~\ref{fig:pmq}).

For the SMDS probe, intro and beyond-intro students were indistinguishable in the rates at which each group applied point and set reasoning. Both intro and beyond-intro students primarily displayed set reasoning in their explanations (67\% and 77\%, respectively). These explanations tended to argue that the data distribution with a narrower spread was the better result. For example, a beyond-intro student wrote, \textit{``The standard deviation and error bars will be smaller for group A.''} Similarly, an intro student argued, \textit{``Although the average is the same, the lower range of the first group is superior.''}

Fewer students applied point reasoning in their responses (24\% of intro students and 17\% of beyond-intro students). Some of these students concluded that both sets of data were equally good because their means were identical, for example \textit{``Averages are how we determine the accuracy of things in physics. The standard deviation/ uncertainty of group B may be larger than for group A, but that doesnt make their result any `worse' than group A''} (beyond-intro student). Another line of point reasoning argued that because uncertainty is always a part of measurement, reducing uncertainty is not important: \textit{``Variety is common in physics experiments. It is not always exact. Therefore, if the experiment is done correctly, there should not be any discussion about which is better because variety is common''} (intro student).

For the DMSS probe, however, we observed distinguishable differences in the rates of point and set reasoning for intro and beyond-intro students. Intro students were more likely to use point reasoning in their explanations (61\%) compared to beyond-intro students (37\%). The students who gave point responses applied varying approaches to comparing the mean values of the two distributions but did not discuss the spread in the data. For example, some of these students argued a difference in means was large or small with no clear justification for the judgement: \textit{``I think that they agree because of how close their averages are''} (beyond-intro student). Other students used the percent difference to make a comparison, for example \textit{``Although their averages are not the same, they are fairly close, and the difference is only a very small percent''} (intro student).

Correspondingly, beyond-intro students were more likely to apply set reasoning (58\%) than intro students (25\%). Set reasoning responses tended to use measures of the variability in the data, such as the standard deviation or the spread, to determine whether the two data sets agreed. For example one beyond-intro student wrote, \textit{``Both groups do not have precise data, and it would reason that the confidence interval of both groups would have an overlap of the other group's data values.''} Similarly, an intro student used the standard deviation to conclude that the two data distributions agreed: \textit{``The average values are within a standard deviation of each other.''}

\subsubsection{Sources of uncertainty}

\begin{figure}[t]
    \centering    \includegraphics[width=.48\textwidth]{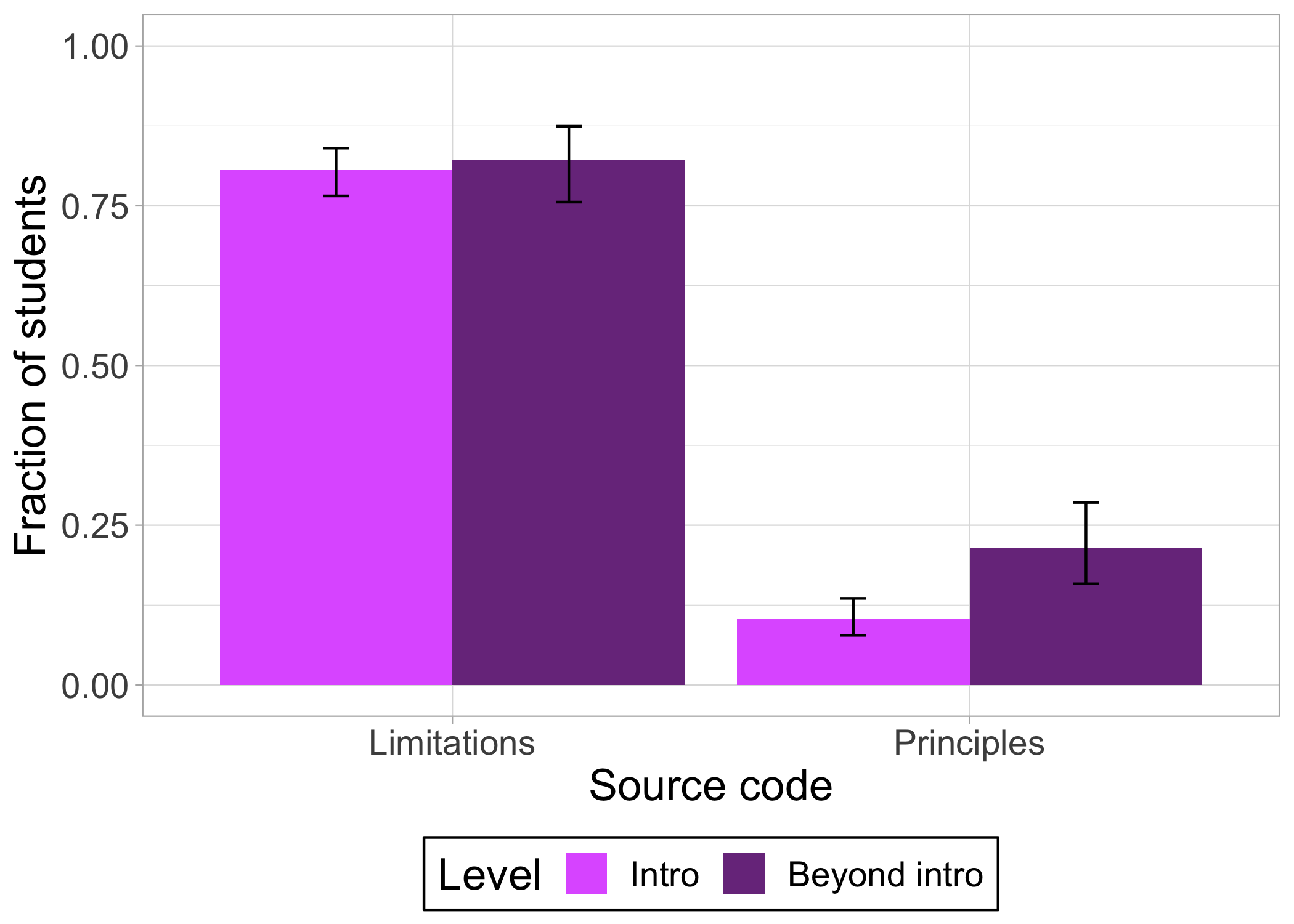}
    \caption{The fraction of intro and beyond-intro students who listed at least one source of uncertainty coded as limitations and principles. Uncertainty bars represent the 95\% confidence interval.}
    \label{fig:sources}
\end{figure}

After the PMQ probes, we then asked students the open-ended Sources question: ``What is causing the shape of the distribution? List as many causes as you can think of.'' We coded student-listed sources of uncertainty as either \textit{limitations} in the experimental apparatus/procedures or as \textit{principles} of the measurement process and here report on the fraction of intro and beyond-intro students who listed at least one source that received each code (see Fig.~\ref{fig:sources}).

The majority of both intro and beyond-intro students listed a source coded as limitations (81\% and 82\%, respectively). These limitations sources included a variety of experimental factors. Some sources were attributed to errors made by the students, such as \textit{``Due to mistakes, there will be some outliers''} (intro student) and \textit{``Error in reading (i.e. the measuring stick was off center, the ball bounced and they are reading the second impact, etc.)''} (beyond-intro student). Other sources highlighted aspects of the setup that would be more difficult for the students to control, such as \textit{``Air currents in the room''} (beyond-intro student), \textit{``Difference of friction between the ball and the ramp caused by blemishes on the ball''} (intro student), and \textit{``Instrumental error''} (beyond-intro student). Overall, the rates at which intro and beyond-intro students identified limitations sources of uncertainty were indistinguishable.

Both intro and beyond-intro students mentioned sources that received the principles code much less frequently than the limitations sources. Furthermore, more beyond-intro students (22\%) than intro students (10\%) listed at least one principles source of uncertainty. These students tended to mention principles sources related to the inherent statistical nature of measurement, for example \textit{``Principle of normal distributions (data shaped like a bell curve)''} (intro student) and \textit{``Overall general Gaussian distribution is due to expected random error''} (beyond-intro student).

\subsubsection{Predictive reasoning}

\begin{figure}
    \centering    \includegraphics[width=.48\textwidth]{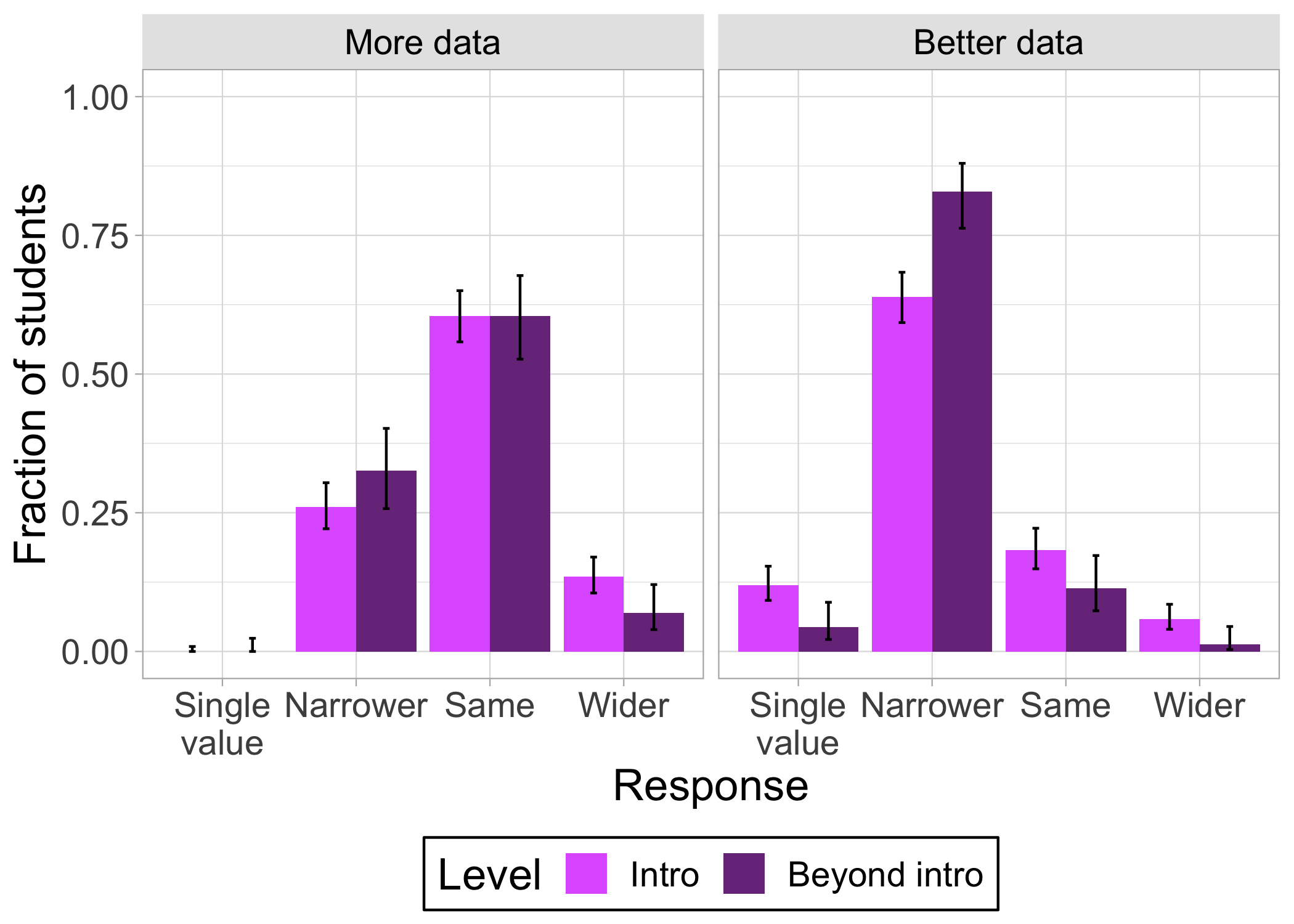}
    \caption{Distribution of intro and beyond-intro students' responses to the more and better data questions. Uncertainty bars represent the 95\% confidence interval.}
    \label{fig:more_better}
\end{figure}

The final set of questions in the survey asked respondents to identify what would happen to the data distribution if 100 more students (More Data) or experts (Better Data) were to conduct the experiment. Respondents were given four choices: a single value is measured (\textit{Single value}), distribution becomes narrower (\textit{Narrower}), distribution stays roughly the same (\textit{Same}), and distribution becomes wider (\textit{Wider}). Here we report on the fraction of intro and beyond-intro students who chose each of these four answer options for the two predictive reasoning questions (see Fig.~\ref{fig:more_better}).

For the More Data question, the fractions of intro and beyond-intro students choosing each option were indistinguishable. A majority of students indicated that the distribution would remain the same if 100 additional students were to collect data (61\% of intro students and 60\% of beyond-intro students), which we consider to be the correct answer. The second most common response for both intro and beyond-intro students was that the distribution would become narrower (26\% of intro students and 33\% of beyond-intro students), which we consider to be an incorrect response.

For the Better Data question, the ordering of answer popularity was the same for intro and beyond-intro students, but the fraction of students who gave each of these responses varied between the two groups. For both groups of students, Narrower was the most common response (64\% of intro students and 83\% of beyond-intro students), followed by Same (18\% of intro students and 11\% of beyond-intro students) and Single value (12\% of intro students and 4\% of beyond-intro students). However, more beyond-intro students answered Narrower (83\%) compared to intro students (64\%), while more intro students answered Single value (12\%) compared to beyond-intro students (4\%), with the differences beyond the 95\% confidence intervals.

\subsubsection{Summary}

Across the five survey questions we analyzed, we observed some instances of similarity between intro and beyond-intro students' answers, while for other questions these two groups answered quite differently. Intro and beyond-intro students primarily applied set reasoning on the SMDS probe, tended to identify limitations sources of uncertainty, and tended to indicate that taking more data would not change the data distribution width. On the other hand, beyond-intro students were more likely to apply set reasoning on the DMSS probe, more likely to list principles sources of uncertainty, and more likely to answer that better data would result in a Narrower distribution (and less likely to answer that a Single value measurement would result) compared to intro students.

\subsection{Possible explanations for differences in responses}

We consider two types of hypotheses for the differences in survey responses between the intro and beyond-intro students and perform appropriate analyses of our data to test them. One possible hypothesis is that the two groups come from different overall populations, characterized by, for example, their majors or differences in the institutions represented at each level. Another possible hypothesis is that the two groups differ only in their educational experience within the physics curriculum. In our analysis, we test each possible hypothesis individually, although we acknowledge that some of the differences we observed may be explained by combinations of variables rather than individual variables.

\subsubsection{Population differences}

We first consider the possibility that variability in responses could be attributable to differences in the populations of the two groups, as distinct from the additional educational physics experiences that the beyond-intro students have had compared to the intro students. 

One population difference relates to the different institutions represented in the samples of intro and beyond-intro students. That is, the data include some institutions that are represented in one group but not the other or that make up different proportions of the sample in each group. Thus, we wanted to confirm that the differences between intro and beyond-intro students were not exclusively explained by institution differences. To do so, we made the same comparisons discussed in Sec.~\ref{sec:overall_results} within a single university (i.e., holding institution constant), Cornell University, as Cornell was the only individual institution where we had large enough sample sizes at both the intro and beyond-intro levels to draw meaningful conclusions. With this subset of the data, we observed the same trends between the intro and beyond-intro students' responses identified in Sec.~\ref{sec:overall_results} (see Fig.~\ref{fig:Cornell} in the appendix). These results are discussed further in Appendix~\ref{sec:figs}. This finding suggests that the differences we observed in the full data set between intro and beyond-intro students were not exclusively explained by the differences in the universities represented in each group. We note also that the Cornell University data do not make up the majority of the full data set at either level, so this analysis is not due to Cornell University driving the trends in the full data set.

The second possibility is that the responses vary due to the different student majors represented in the samples of intro and beyond-intro students. Physics majors comprise 92\% of our beyond-intro sample but only 15\% of our intro sample. We compared the responses of intro-level physics majors to intro-level non-physics majors. For four of the five survey questions (namely SMDS, Sources, More data, and Better data) we observed that intro physics majors and intro non-physics majors responded similarly (see Fig.~\ref{fig:major} in the appendix and Fig.~\ref{fig:major_pmq}). 
\begin{figure}
    \centering
    \includegraphics[width=.48\textwidth]{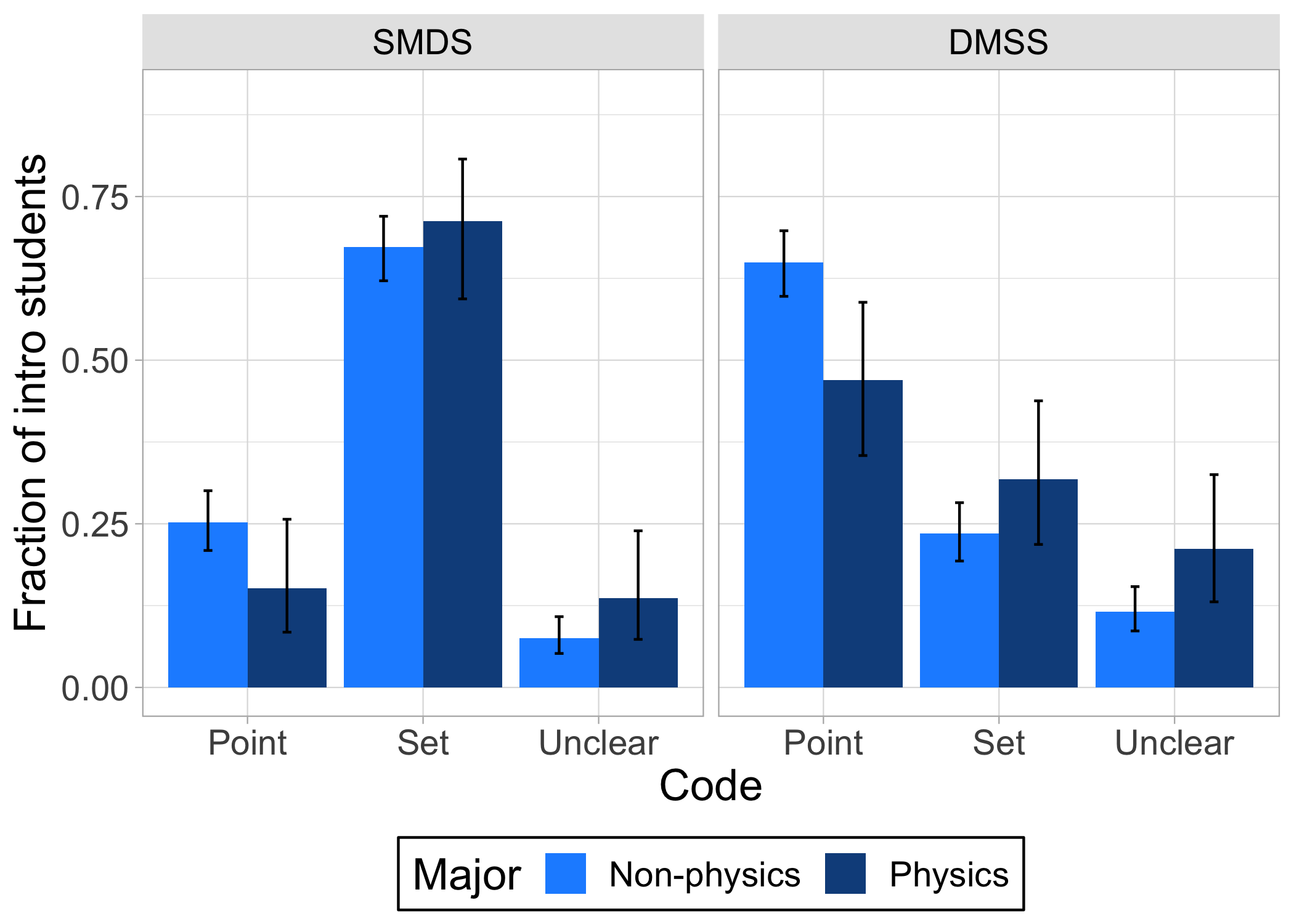}
    \caption{Codes applied to intro physics majors' and intro non-physics majors' responses to the SMDS and DMSS probes. Uncertainty bars represent the 95\% confidence interval.}
    \label{fig:major_pmq}
\end{figure}
For the DMSS probe, we observed a small difference in reasoning based on major. More intro non-physics majors than intro physics majors exhibited point reasoning in their responses (65\% and 50\%, respectively; see Fig.~\ref{fig:major_pmq}). However, the fractions of non-physics majors and physics majors who used set reasoning were indistinguishable (23\% and 32\%, respectively). These results indicate that the differences in majors between the intro and beyond-intro students may partly explain the lower rate of point reasoning in beyond-intro students' responses but cannot explain the higher rate of set reasoning in beyond-intro students' responses.

\subsubsection{Educational experiences}

The above results indicate that population differences based on institution or major cannot fully explain the observed differences in intro and beyond-intro student reasoning. The differences in reasoning, therefore, are likely also due to differences in physics educational experiences between the intro and beyond-intro students. Here we consider two types of educational experiences that may impact student reasoning about uncertainty: lab courses and research experience.

One of the places we would expect beyond-intro students to learn more about uncertainty is in lab courses taken beyond the intro level. To test this explanation, we compared beyond-intro students' responses to the five survey questions based on whether they had taken only intro lab courses or had taken (or were currently taking) at least one lab course beyond the intro level. We found no differences in student responses to any of the survey questions based on whether students had taken a beyond-intro lab course (see Fig.~\ref{fig:lab} in the appendix).

Another environment in which we might expect beyond-intro students to learn more about uncertainty is conducting research in an experimental laboratory. To test this explanation, we compared beyond-intro students' responses to the five survey questions based on whether they had research experience in an experimental context. We found no differences in student responses to any of the survey questions based on research experience (see Fig.~\ref{fig:research} in the appendix). Overall, we found no evidence that either lab courses taken or research experience could explain the differences in intro and beyond-intro students' responses.

\section{Discussion}

In this study, we probed intro and beyond-intro students' ideas about uncertainty using five survey questions related to procedural reasoning, sources of uncertainty, and predictive reasoning. We found that intro and beyond-intro students gave similar answers for the SMDS and More Data questions and in identifying limitations sources of uncertainty but different answers for the DMSS and Better Data questions and in identifying principles sources of uncertainty.

We found that intro and beyond-intro students answered similarly on questions where both groups were mostly using expertlike thinking. Both intro (67\%) and beyond-intro (77\%) students primarily used set reasoning in their responses to the SMDS probe. This result aligns with previously reported rates of set thinking on the SMDS probe after intro lab instruction~\footnote{Other work has also looked at student responses to both the SMDS and DMSS probes but either collapsed results for the SMDS and DMSS probes into a single category of Data Comparison~\cite{Lubben2001,Buffler2001}  or did not report the frequency of point/set codes at the student level~\cite{Pollard2020}. As a result, we cannot compare our results for individual probes directly to these previous findings.}. For example, Pillay \textit{et al.}~\cite{Pillay2008} found that 64\% of intro students at the University of Cape Town used set reasoning on this probe after taking a lab course designed to help students develop set thinking. Similarly, Wilson \textit{et al.}~\cite{Wilson2022} reported that approximately 70\% of responses in a sample of mixed pre and post surveys from the University of Colorado Boulder included set reasoning. For the More Data question, intro and beyond-intro students also answered similarly, with most students giving the correct answer of ``same'' (61\% and 60\%, respectively). The high rates of expertlike reasoning among intro-level students may explain why we observed no difference between intro and beyond-intro students' responses: most intro students have mastered the relevant ideas about uncertainty in the context of these questions, so there is limited room for improvement from intro to beyond-intro levels.
 

Another similarity in intro and beyond-intro students' responses was in identifying limitations sources of uncertainty. The majority of both intro (81\%) and beyond-intro (82\%) students listed at least one limitation in the procedures or physical setup of the experiment when asked the Sources question. This is unsurprising, as prior work has found that intro students are able to identify a wide variety of sources of uncertainty in a lab setting~\cite{Holmes2015,Etkina2008,Sere1993}. Unlike the more data and SMDS items discussed above, we cannot characterize students listing limitations sources as expertlike or novicelike. Students can list a wide variety of limitations in an experiment, ranging from actionable or quantifiable sources of uncertainty, such as varying force applied while dropping a ball or the instrumental precision of a ruler, to the more vague and unproductive ``human error''~\cite{Hu2018,Holmes2015,Sere2001,Evangelinos2002,Allie2003}. Furthermore, students often struggle to quantify uncertainty associated with limitations in their experiment~\cite{Sere1993,Etkina2008}, which means listing limitations alone does not demonstrate expertise. Future work should disentangle different productive and unproductive modes of reasoning about limitations in experiments.

For the DMSS probe and Better Data question, in contrast, we observed that beyond-intro students exhibited more expertlike reasoning than intro students. For the DMSS probe, our results indicate low levels of set reasoning among intro students (25\%), which aligns with the pre-instruction rates of set reasoning reported in prior work~\cite{Pillay2008,Allie1998}. Notably, rates of set reasoning were much higher in Pillay \textit{et al.}'s~\cite{Pillay2008} post-instruction survey (75\%) and in Wilson \textit{et al.}'s mixed pre/post data set (approximately 60\%) than in our data. This contradiction may suggest that the intro lab courses in our dataset may be less effective in teaching set reasoning compared to the lab courses in Pillay \textit{et al.}'s and Wilson \textit{et al.}'s studies. In spite of this apparent shortcoming in intro lab courses, however, the majority of beyond-intro students in our study exhibited set reasoning on the DMSS probe (58\%), though still less frequent than in Pillay \textit{et al.}'s study.

For the Better Data question, we observed that beyond-intro students (83\%) were more likely than intro students (64\%) to indicate that experts would measure a Narrower distribution, corresponding to a lower fraction of beyond-intro students (4\%) than intro students (12\%) who indicated that experts would measure a single value. Believing that experts would measure a single value is aligned with point reasoning, implying that all uncertainty in an experiment can be eliminated and that a single measurement can produce the ``true value''~\cite{Allie1998,Buffler2001,Lubben2001,Leach1998,Evangelinos1999}. Although this response was fairly uncommon for both intro and beyond-intro students, the lower fraction of beyond-intro students responding ``single value'' suggests that beyond-intro educational experiences may be effective for eliminating this view of uncertainty from students' understanding of measurement.

One aspect of student reasoning that may help explain the differences in the DMSS probe and Better Data question is students' ideas about sources of uncertainty. Although rare, more beyond-intro students (22\%) than intro students (10\%) described uncertainty as a principle inherent to the measurement process. Given that previous work has argued that holding this conception of uncertainty can help students apply set reasoning~\cite{Evangelinos1999,Evangelinos2002,Allie2003,Pillay2008,Buffler2008}, this may help explain why more beyond-intro than intro students exhibited set reasoning on the DMSS probe and why fewer beyond-intro than intro students answered Single value on the Better Data question. Additional research is necessary to understand how reasoning about sources of uncertainty is connected to point and set reasoning on specific PMQ probes and predictive reasoning questions.

We attempted to determine what educational experiences might explain the differences between intro and beyond-intro students' responses on the survey. We tested whether taking at least one lab course beyond the introductory level or having experimental research experience could explain the differences in student reasoning. We found no evidence, however, that either of these educational experiences alone could explain the differences in intro and beyond-intro students' reasoning. Overall, our results suggest that students' educational experiences beyond the intro level may be enhancing students' reasoning about some aspects of measurement, such as considering uncertainty when comparing two data sets and recognizing that uncertainty is a fundamental aspect of experimental measurement and cannot be eliminated. However, current beyond-intro educational experiences may be less effective for changing other aspects of students' reasoning about uncertainty, such as teaching students that smaller spread in data is desirable or that collecting more data will not change the shape of a data distribution. More research is necessary to identify what specific aspects of beyond-intro students' educational experiences are effective for shifting students' reasoning about uncertainty and evaluate how interventions and course transformations can be used to improve lab instruction related to uncertainty beyond the intro level. In the future, we intend to administer this survey pre and post to students in a variety of physics courses to better understand how specific pedagogical practices impact student reasoning about uncertainty.
 
\acknowledgments{This material is based upon work supported by the National Science Foundation Graduate Research Fellowship under Grant No.~DGE-2139899 and National Science Foundation Grants No.~DUE-1808945 and No.~DUE-1809178. We thank Courtney White, Matthew Dew, and Andy Schang for their contributions to our previous investigations of student reasoning about uncertainty.}

\bibliography{_refs} 

\begin{thebibliography}{38}%
\makeatletter
\providecommand \@ifxundefined [1]{%
 \@ifx{#1\undefined}
}%
\providecommand \@ifnum [1]{%
 \ifnum #1\expandafter \@firstoftwo
 \else \expandafter \@secondoftwo
 \fi
}%
\providecommand \@ifx [1]{%
 \ifx #1\expandafter \@firstoftwo
 \else \expandafter \@secondoftwo
 \fi
}%
\providecommand \natexlab [1]{#1}%
\providecommand \enquote  [1]{``#1''}%
\providecommand \bibnamefont  [1]{#1}%
\providecommand \bibfnamefont [1]{#1}%
\providecommand \citenamefont [1]{#1}%
\providecommand \href@noop [0]{\@secondoftwo}%
\providecommand \href [0]{\begingroup \@sanitize@url \@href}%
\providecommand \@href[1]{\@@startlink{#1}\@@href}%
\providecommand \@@href[1]{\endgroup#1\@@endlink}%
\providecommand \@sanitize@url [0]{\catcode `\\12\catcode `\$12\catcode
  `\&12\catcode `\#12\catcode `\^12\catcode `\_12\catcode `\%12\relax}%
\providecommand \@@startlink[1]{}%
\providecommand \@@endlink[0]{}%
\providecommand \url  [0]{\begingroup\@sanitize@url \@url }%
\providecommand \@url [1]{\endgroup\@href {#1}{\urlprefix }}%
\providecommand \urlprefix  [0]{URL }%
\providecommand \Eprint [0]{\href }%
\providecommand \doibase [0]{https://doi.org/}%
\providecommand \selectlanguage [0]{\@gobble}%
\providecommand \bibinfo  [0]{\@secondoftwo}%
\providecommand \bibfield  [0]{\@secondoftwo}%
\providecommand \translation [1]{[#1]}%
\providecommand \BibitemOpen [0]{}%
\providecommand \bibitemStop [0]{}%
\providecommand \bibitemNoStop [0]{.\EOS\space}%
\providecommand \EOS [0]{\spacefactor3000\relax}%
\providecommand \BibitemShut  [1]{\csname bibitem#1\endcsname}%
\let\auto@bib@innerbib\@empty
\bibitem [{\citenamefont {Kozminski}\ \emph {et~al.}(2014)\citenamefont
  {Kozminski}, \citenamefont {Beverly}, \citenamefont {Deardorff},
  \citenamefont {Dietz}, \citenamefont {Eblen-Zayas}, \citenamefont {Hobbs},
  \citenamefont {Lewandowski}, \citenamefont {Lindaas}, \citenamefont {Reagan},
  \citenamefont {Tagg}, \citenamefont {Williams},\ and\ \citenamefont
  {Zwickl}}]{Kozminski2014}%
  \BibitemOpen
  \bibfield  {author} {\bibinfo {author} {\bibfnamefont {J.}~\bibnamefont
  {Kozminski}}, \bibinfo {author} {\bibfnamefont {N.}~\bibnamefont {Beverly}},
  \bibinfo {author} {\bibfnamefont {D.}~\bibnamefont {Deardorff}}, \bibinfo
  {author} {\bibfnamefont {R.}~\bibnamefont {Dietz}}, \bibinfo {author}
  {\bibfnamefont {M.}~\bibnamefont {Eblen-Zayas}}, \bibinfo {author}
  {\bibfnamefont {R.}~\bibnamefont {Hobbs}}, \bibinfo {author} {\bibfnamefont
  {H.}~\bibnamefont {Lewandowski}}, \bibinfo {author} {\bibfnamefont
  {S.}~\bibnamefont {Lindaas}}, \bibinfo {author} {\bibfnamefont
  {A.}~\bibnamefont {Reagan}}, \bibinfo {author} {\bibfnamefont
  {R.}~\bibnamefont {Tagg}}, \bibinfo {author} {\bibfnamefont {J.}~\bibnamefont
  {Williams}},\ and\ \bibinfo {author} {\bibfnamefont {B.}~\bibnamefont
  {Zwickl}},\ }\href
  {https://www.aapt.org/resources/upload/labguidlinesdocument_ebendorsed_nov10.pdf}
  {\emph {\bibinfo {title} {{AAPT} recommendations for the undergraduate
  physics laboratory curriculum}}},\ \bibinfo {type} {Tech. Rep.}\ (\bibinfo
  {year} {2014})\BibitemShut {NoStop}%
\bibitem [{\citenamefont {Walsh}\ \emph {et~al.}(2019)\citenamefont {Walsh},
  \citenamefont {Quinn}, \citenamefont {Wieman},\ and\ \citenamefont
  {Holmes}}]{Walsh2019}%
  \BibitemOpen
  \bibfield  {author} {\bibinfo {author} {\bibfnamefont {C.}~\bibnamefont
  {Walsh}}, \bibinfo {author} {\bibfnamefont {K.~N.}\ \bibnamefont {Quinn}},
  \bibinfo {author} {\bibfnamefont {C.}~\bibnamefont {Wieman}},\ and\ \bibinfo
  {author} {\bibfnamefont {N.~G.}\ \bibnamefont {Holmes}},\ }\bibfield  {title}
  {\bibinfo {title} {Quantifying critical thinking: Development and validation
  of the physics lab inventory of critical thinking},\ }\href
  {https://doi.org/10.1103/PhysRevPhysEducRes.15.010135} {\bibfield  {journal}
  {\bibinfo  {journal} {Phys. Rev. Phys. Educ. Res.}\ }\textbf {\bibinfo
  {volume} {15}},\ \bibinfo {pages} {010135} (\bibinfo {year}
  {2019})}\BibitemShut {NoStop}%
\bibitem [{\citenamefont {Holmes}\ and\ \citenamefont
  {Smith}(2019)}]{Holmes2019}%
  \BibitemOpen
  \bibfield  {author} {\bibinfo {author} {\bibfnamefont {N.~G.}\ \bibnamefont
  {Holmes}}\ and\ \bibinfo {author} {\bibfnamefont {E.~M.}\ \bibnamefont
  {Smith}},\ }\bibfield  {title} {\bibinfo {title} {Operationalizing the {AAPT}
  learning goals for the lab},\ }\href {https://doi.org/10.1119/1.5098916}
  {\bibfield  {journal} {\bibinfo  {journal} {The Physics Teacher}\ }\textbf
  {\bibinfo {volume} {57}},\ \bibinfo {pages} {296} (\bibinfo {year}
  {2019})}\BibitemShut {NoStop}%
\bibitem [{\citenamefont {Holmes}\ \emph {et~al.}(2015)\citenamefont {Holmes},
  \citenamefont {Wieman},\ and\ \citenamefont {Bonn}}]{Holmes2015a}%
  \BibitemOpen
  \bibfield  {author} {\bibinfo {author} {\bibfnamefont {N.~G.}\ \bibnamefont
  {Holmes}}, \bibinfo {author} {\bibfnamefont {C.~E.}\ \bibnamefont {Wieman}},\
  and\ \bibinfo {author} {\bibfnamefont {D.~A.}\ \bibnamefont {Bonn}},\
  }\bibfield  {title} {\bibinfo {title} {Teaching critical thinking},\ }\href
  {https://doi.org/10.1073/pnas.1505329112} {\bibfield  {journal} {\bibinfo
  {journal} {Proceedings of the National Academy of Sciences}\ }\textbf
  {\bibinfo {volume} {112}},\ \bibinfo {pages} {11199} (\bibinfo {year}
  {2015})}\BibitemShut {NoStop}%
\bibitem [{\citenamefont {Holmes}\ and\ \citenamefont
  {Wieman}(2015)}]{Holmes2015}%
  \BibitemOpen
  \bibfield  {author} {\bibinfo {author} {\bibfnamefont {N.~G.}\ \bibnamefont
  {Holmes}}\ and\ \bibinfo {author} {\bibfnamefont {C.}~\bibnamefont
  {Wieman}},\ }\bibfield  {title} {\bibinfo {title} {Assessing modeling in the
  lab: Uncertainty and measurement},\ }in\ \href@noop {} {\emph {\bibinfo
  {booktitle} {2015 Conference on Laboratory Instruction Beyond the First Year
  of College}}},\ \bibinfo {series and number} {BFY Conference}\ (\bibinfo
  {address} {College Park, MD},\ \bibinfo {year} {2015})\ pp.\ \bibinfo {pages}
  {44--47}\BibitemShut {NoStop}%
\bibitem [{\citenamefont {Allie}\ \emph {et~al.}(2003)\citenamefont {Allie},
  \citenamefont {Buffler}, \citenamefont {Campbell}, \citenamefont {Lubben},
  \citenamefont {Evangelinos}, \citenamefont {Psillos},\ and\ \citenamefont
  {Valassiades}}]{Allie2003}%
  \BibitemOpen
  \bibfield  {author} {\bibinfo {author} {\bibfnamefont {S.}~\bibnamefont
  {Allie}}, \bibinfo {author} {\bibfnamefont {A.}~\bibnamefont {Buffler}},
  \bibinfo {author} {\bibfnamefont {B.}~\bibnamefont {Campbell}}, \bibinfo
  {author} {\bibfnamefont {F.}~\bibnamefont {Lubben}}, \bibinfo {author}
  {\bibfnamefont {D.}~\bibnamefont {Evangelinos}}, \bibinfo {author}
  {\bibfnamefont {D.}~\bibnamefont {Psillos}},\ and\ \bibinfo {author}
  {\bibfnamefont {O.}~\bibnamefont {Valassiades}},\ }\bibfield  {title}
  {\bibinfo {title} {Teaching measurement in the introductory physics
  laboratory},\ }\href {https://doi.org/10.1119/1.1616479} {\bibfield
  {journal} {\bibinfo  {journal} {The Physics Teacher}\ }\textbf {\bibinfo
  {volume} {41}},\ \bibinfo {pages} {394} (\bibinfo {year} {2003})}\BibitemShut
  {NoStop}%
\bibitem [{\citenamefont {Kung}(2005)}]{Kung2005}%
  \BibitemOpen
  \bibfield  {author} {\bibinfo {author} {\bibfnamefont {R.~L.}\ \bibnamefont
  {Kung}},\ }\bibfield  {title} {\bibinfo {title} {Teaching the concepts of
  measurement: An example of a concept-based laboratory course},\ }\href
  {https://doi.org/10.1119/1.1881253} {\bibfield  {journal} {\bibinfo
  {journal} {American Journal of Physics}\ }\textbf {\bibinfo {volume} {73}},\
  \bibinfo {pages} {771} (\bibinfo {year} {2005})}\BibitemShut {NoStop}%
\bibitem [{\citenamefont {Buffler}\ \emph {et~al.}(2008)\citenamefont
  {Buffler}, \citenamefont {Allie},\ and\ \citenamefont
  {Lubben}}]{Buffler2008}%
  \BibitemOpen
  \bibfield  {author} {\bibinfo {author} {\bibfnamefont {A.}~\bibnamefont
  {Buffler}}, \bibinfo {author} {\bibfnamefont {S.}~\bibnamefont {Allie}},\
  and\ \bibinfo {author} {\bibfnamefont {F.}~\bibnamefont {Lubben}},\
  }\bibfield  {title} {\bibinfo {title} {Teaching measurement and uncertainty
  the {GUM} way},\ }\href {https://doi.org/10.1119/1.3023655} {\bibfield
  {journal} {\bibinfo  {journal} {The Physics Teacher}\ }\textbf {\bibinfo
  {volume} {46}},\ \bibinfo {pages} {539} (\bibinfo {year} {2008})}\BibitemShut
  {NoStop}%
\bibitem [{\citenamefont {S\'{e}r\'{e}}\ \emph {et~al.}(1993)\citenamefont
  {S\'{e}r\'{e}}, \citenamefont {Journeaux},\ and\ \citenamefont
  {Larcher}}]{Sere1993}%
  \BibitemOpen
  \bibfield  {author} {\bibinfo {author} {\bibfnamefont {M.-G.}\ \bibnamefont
  {S\'{e}r\'{e}}}, \bibinfo {author} {\bibfnamefont {R.}~\bibnamefont
  {Journeaux}},\ and\ \bibinfo {author} {\bibfnamefont {C.}~\bibnamefont
  {Larcher}},\ }\bibfield  {title} {\bibinfo {title} {Learning the statistical
  analysis of measurement errors},\ }\href
  {https://doi.org/10.1080/0950069930150406} {\bibfield  {journal} {\bibinfo
  {journal} {International Journal of Science Education}\ }\textbf {\bibinfo
  {volume} {15}},\ \bibinfo {pages} {427} (\bibinfo {year} {1993})}\BibitemShut
  {NoStop}%
\bibitem [{\citenamefont {Coelho}\ and\ \citenamefont
  {S\'{e}r\'{e}}(1998)}]{Coelho1998}%
  \BibitemOpen
  \bibfield  {author} {\bibinfo {author} {\bibfnamefont {S.~M.}\ \bibnamefont
  {Coelho}}\ and\ \bibinfo {author} {\bibfnamefont {M.-G.}\ \bibnamefont
  {S\'{e}r\'{e}}},\ }\bibfield  {title} {\bibinfo {title} {Pupils' reasoning
  and practice during hands-on activities in the measurement phase},\ }\href
  {https://doi.org/10.1080/0263514980160107} {\bibfield  {journal} {\bibinfo
  {journal} {Research in Science \& Technological Education}\ }\textbf
  {\bibinfo {volume} {16}},\ \bibinfo {pages} {79} (\bibinfo {year}
  {1998})}\BibitemShut {NoStop}%
\bibitem [{\citenamefont {Leach}\ \emph {et~al.}(1998)\citenamefont {Leach},
  \citenamefont {Millar}, \citenamefont {Ryder}, \citenamefont {S\'{e}r\'{e}},
  \citenamefont {Hammelev}, \citenamefont {Niedderer},\ and\ \citenamefont
  {Tselfes}}]{Leach1998}%
  \BibitemOpen
  \bibfield  {author} {\bibinfo {author} {\bibfnamefont {J.}~\bibnamefont
  {Leach}}, \bibinfo {author} {\bibfnamefont {R.}~\bibnamefont {Millar}},
  \bibinfo {author} {\bibfnamefont {J.}~\bibnamefont {Ryder}}, \bibinfo
  {author} {\bibfnamefont {M.-G.}\ \bibnamefont {S\'{e}r\'{e}}}, \bibinfo
  {author} {\bibfnamefont {D.}~\bibnamefont {Hammelev}}, \bibinfo {author}
  {\bibfnamefont {H.}~\bibnamefont {Niedderer}},\ and\ \bibinfo {author}
  {\bibfnamefont {V.}~\bibnamefont {Tselfes}},\ }\href@noop {} {\emph {\bibinfo
  {title} {Survey 2: Students' Images of Science as They Relate to Labwork
  Learning}}},\ \bibinfo {type} {Tech. Rep.}\ \bibinfo {number} {Working Paper
  6}\ (\bibinfo  {institution} {Targeted Socio-Economic Research Programme
  Project PL 95-2005.},\ \bibinfo {year} {1998})\BibitemShut {NoStop}%
\bibitem [{\citenamefont {Evangelinos}\ \emph {et~al.}(1999)\citenamefont
  {Evangelinos}, \citenamefont {Valassiades},\ and\ \citenamefont
  {Psillos}}]{Evangelinos1999}%
  \BibitemOpen
  \bibfield  {author} {\bibinfo {author} {\bibfnamefont {D.}~\bibnamefont
  {Evangelinos}}, \bibinfo {author} {\bibfnamefont {O.}~\bibnamefont
  {Valassiades}},\ and\ \bibinfo {author} {\bibfnamefont {D.}~\bibnamefont
  {Psillos}},\ }\bibfield  {title} {\bibinfo {title} {Undergraduate students'
  views about the approximate nature of measurement results},\ }in\ \href
  {https://www.researchgate.net/publication/349075786_Undergraduate_students'_views_about_the_approximate_nature_of_measurement_results}
  {\emph {\bibinfo {booktitle} {Proceedings of the 2nd Second International
  Conference of the European Science Education Research Association}}}\
  (\bibinfo {year} {1999})\BibitemShut {NoStop}%
\bibitem [{\citenamefont {Allie}\ \emph {et~al.}(1998)\citenamefont {Allie},
  \citenamefont {Buffler}, \citenamefont {Campbell},\ and\ \citenamefont
  {Lubben}}]{Allie1998}%
  \BibitemOpen
  \bibfield  {author} {\bibinfo {author} {\bibfnamefont {S.}~\bibnamefont
  {Allie}}, \bibinfo {author} {\bibfnamefont {A.}~\bibnamefont {Buffler}},
  \bibinfo {author} {\bibfnamefont {B.}~\bibnamefont {Campbell}},\ and\
  \bibinfo {author} {\bibfnamefont {F.}~\bibnamefont {Lubben}},\ }\bibfield
  {title} {\bibinfo {title} {First-year physics students' perceptions of the
  quality of experimental measurements},\ }\href
  {https://doi.org/10.1080/0950069980200405} {\bibfield  {journal} {\bibinfo
  {journal} {International Journal of Science Education}\ }\textbf {\bibinfo
  {volume} {20}},\ \bibinfo {pages} {447} (\bibinfo {year} {1998})}\BibitemShut
  {NoStop}%
\bibitem [{\citenamefont {Buffler}\ \emph {et~al.}(2001)\citenamefont
  {Buffler}, \citenamefont {Allie},\ and\ \citenamefont
  {Lubben}}]{Buffler2001}%
  \BibitemOpen
  \bibfield  {author} {\bibinfo {author} {\bibfnamefont {A.}~\bibnamefont
  {Buffler}}, \bibinfo {author} {\bibfnamefont {S.}~\bibnamefont {Allie}},\
  and\ \bibinfo {author} {\bibfnamefont {F.}~\bibnamefont {Lubben}},\
  }\bibfield  {title} {\bibinfo {title} {The development of first year physics
  students' ideas about measurement in terms of point and set paradigms},\
  }\href {https://doi.org/10.1080/09500690110039567} {\bibfield  {journal}
  {\bibinfo  {journal} {International Journal of Science Education}\ }\textbf
  {\bibinfo {volume} {23}},\ \bibinfo {pages} {1137} (\bibinfo {year}
  {2001})}\BibitemShut {NoStop}%
\bibitem [{\citenamefont {Lubben}\ \emph {et~al.}(2001)\citenamefont {Lubben},
  \citenamefont {Campbell}, \citenamefont {Buffler},\ and\ \citenamefont
  {Allie}}]{Lubben2001}%
  \BibitemOpen
  \bibfield  {author} {\bibinfo {author} {\bibfnamefont {F.}~\bibnamefont
  {Lubben}}, \bibinfo {author} {\bibfnamefont {B.}~\bibnamefont {Campbell}},
  \bibinfo {author} {\bibfnamefont {A.}~\bibnamefont {Buffler}},\ and\ \bibinfo
  {author} {\bibfnamefont {S.}~\bibnamefont {Allie}},\ }\bibfield  {title}
  {\bibinfo {title} {Point and set reasoning in practical science measurement
  by entering university freshmen},\ }\href
  {https://doi.org/https://doi.org/10.1002/sce.1012} {\bibfield  {journal}
  {\bibinfo  {journal} {Science Education}\ }\textbf {\bibinfo {volume} {85}},\
  \bibinfo {pages} {311} (\bibinfo {year} {2001})}\BibitemShut {NoStop}%
\bibitem [{\citenamefont {Volkwyn}\ \emph {et~al.}(2008)\citenamefont
  {Volkwyn}, \citenamefont {Allie}, \citenamefont {Buffler},\ and\
  \citenamefont {Lubben}}]{Volkwyn2008}%
  \BibitemOpen
  \bibfield  {author} {\bibinfo {author} {\bibfnamefont {T.~S.}\ \bibnamefont
  {Volkwyn}}, \bibinfo {author} {\bibfnamefont {S.}~\bibnamefont {Allie}},
  \bibinfo {author} {\bibfnamefont {A.}~\bibnamefont {Buffler}},\ and\ \bibinfo
  {author} {\bibfnamefont {F.}~\bibnamefont {Lubben}},\ }\bibfield  {title}
  {\bibinfo {title} {Impact of a conventional introductory laboratory course on
  the understanding of measurement},\ }\href
  {https://doi.org/10.1103/PhysRevSTPER.4.010108} {\bibfield  {journal}
  {\bibinfo  {journal} {Phys. Rev. ST Phys. Educ. Res.}\ }\textbf {\bibinfo
  {volume} {4}},\ \bibinfo {pages} {010108} (\bibinfo {year}
  {2008})}\BibitemShut {NoStop}%
\bibitem [{\citenamefont {Wan}(2022)}]{Wan2022}%
  \BibitemOpen
  \bibfield  {author} {\bibinfo {author} {\bibfnamefont {T.}~\bibnamefont
  {Wan}},\ }\bibfield  {title} {\bibinfo {title} {Investigating student
  reasoning about measurement uncertainty and ability to draw conclusions from
  measurement data in inquiry-based university physics labs},\ }\href
  {https://doi.org/10.1080/09500693.2022.2156824} {\bibfield  {journal}
  {\bibinfo  {journal} {International Journal of Science Education}\ }\textbf
  {\bibinfo {volume} {0}},\ \bibinfo {pages} {1} (\bibinfo {year}
  {2022})}\BibitemShut {NoStop}%
\bibitem [{\citenamefont {Pollard}\ \emph {et~al.}(2020)\citenamefont
  {Pollard}, \citenamefont {Werth}, \citenamefont {Hobbs},\ and\ \citenamefont
  {Lewandowski}}]{Pollard2020}%
  \BibitemOpen
  \bibfield  {author} {\bibinfo {author} {\bibfnamefont {B.}~\bibnamefont
  {Pollard}}, \bibinfo {author} {\bibfnamefont {A.}~\bibnamefont {Werth}},
  \bibinfo {author} {\bibfnamefont {R.}~\bibnamefont {Hobbs}},\ and\ \bibinfo
  {author} {\bibfnamefont {H.~J.}\ \bibnamefont {Lewandowski}},\ }\bibfield
  {title} {\bibinfo {title} {Impact of a course transformation on students'
  reasoning about measurement uncertainty},\ }\href
  {https://doi.org/10.1103/PhysRevPhysEducRes.16.020160} {\bibfield  {journal}
  {\bibinfo  {journal} {Phys. Rev. Phys. Educ. Res.}\ }\textbf {\bibinfo
  {volume} {16}},\ \bibinfo {pages} {020160} (\bibinfo {year}
  {2020})}\BibitemShut {NoStop}%
\bibitem [{\citenamefont {Kung}\ and\ \citenamefont {Linder}(2006)}]{Kung2006}%
  \BibitemOpen
  \bibfield  {author} {\bibinfo {author} {\bibfnamefont {R.~L.}\ \bibnamefont
  {Kung}}\ and\ \bibinfo {author} {\bibfnamefont {C.}~\bibnamefont {Linder}},\
  }\bibfield  {title} {\bibinfo {title} {University students' ideas about data
  processing and data comparison in a physics laboratory course},\ }\href
  {https://doi.org/https://doi.org/10.5617/nordina.423} {\bibfield  {journal}
  {\bibinfo  {journal} {Nordina}\ }\textbf {\bibinfo {volume} {2}},\ \bibinfo
  {pages} {40} (\bibinfo {year} {2006})}\BibitemShut {NoStop}%
\bibitem [{\citenamefont {Pillay}\ \emph {et~al.}(2008)\citenamefont {Pillay},
  \citenamefont {Buffler}, \citenamefont {Lubben},\ and\ \citenamefont
  {Allie}}]{Pillay2008}%
  \BibitemOpen
  \bibfield  {author} {\bibinfo {author} {\bibfnamefont {S.}~\bibnamefont
  {Pillay}}, \bibinfo {author} {\bibfnamefont {A.}~\bibnamefont {Buffler}},
  \bibinfo {author} {\bibfnamefont {F.}~\bibnamefont {Lubben}},\ and\ \bibinfo
  {author} {\bibfnamefont {S.}~\bibnamefont {Allie}},\ }\bibfield  {title}
  {\bibinfo {title} {Effectiveness of a {GUM-compliant} course for teaching
  measurement in the introductory physics lab},\ }\href
  {https://iopscience.iop.org/article/10.1088/0143-0807/29/3/024} {\bibfield
  {journal} {\bibinfo  {journal} {European Journal of Physics}\ }\textbf
  {\bibinfo {volume} {29}},\ \bibinfo {pages} {647} (\bibinfo {year}
  {2008})}\BibitemShut {NoStop}%
\bibitem [{\citenamefont {Etkina}\ \emph {et~al.}(2008)\citenamefont {Etkina},
  \citenamefont {Karelina},\ and\ \citenamefont
  {Ruibal-Villasenor}}]{Etkina2008}%
  \BibitemOpen
  \bibfield  {author} {\bibinfo {author} {\bibfnamefont {E.}~\bibnamefont
  {Etkina}}, \bibinfo {author} {\bibfnamefont {A.}~\bibnamefont {Karelina}},\
  and\ \bibinfo {author} {\bibfnamefont {M.}~\bibnamefont
  {Ruibal-Villasenor}},\ }\bibfield  {title} {\bibinfo {title} {How long does
  it take? a study of student acquisition of scientific abilities},\ }\href
  {https://doi.org/10.1103/PhysRevSTPER.4.020108} {\bibfield  {journal}
  {\bibinfo  {journal} {Phys. Rev. ST Phys. Educ. Res.}\ }\textbf {\bibinfo
  {volume} {4}},\ \bibinfo {pages} {020108} (\bibinfo {year}
  {2008})}\BibitemShut {NoStop}%
\bibitem [{\citenamefont {Hu}\ and\ \citenamefont {Zwickl}(2018)}]{Hu2018}%
  \BibitemOpen
  \bibfield  {author} {\bibinfo {author} {\bibfnamefont {D.}~\bibnamefont
  {Hu}}\ and\ \bibinfo {author} {\bibfnamefont {B.~M.}\ \bibnamefont
  {Zwickl}},\ }\bibfield  {title} {\bibinfo {title} {Examining students' views
  about validity of experiments: From introductory to {Ph.D.} students},\
  }\href {https://doi.org/10.1103/PhysRevPhysEducRes.14.010121} {\bibfield
  {journal} {\bibinfo  {journal} {Phys. Rev. Phys. Educ. Res.}\ }\textbf
  {\bibinfo {volume} {14}},\ \bibinfo {pages} {010121} (\bibinfo {year}
  {2018})}\BibitemShut {NoStop}%
\bibitem [{\citenamefont {S\'{e}r\'{e}}\ \emph {et~al.}(2001)\citenamefont
  {S\'{e}r\'{e}}, \citenamefont {Fernandez-Gonzalez}, \citenamefont {Gallegos},
  \citenamefont {Manuel}, \citenamefont {Perales},\ and\ \citenamefont
  {Leach}}]{Sere2001}%
  \BibitemOpen
  \bibfield  {author} {\bibinfo {author} {\bibfnamefont {M.-G.}\ \bibnamefont
  {S\'{e}r\'{e}}}, \bibinfo {author} {\bibfnamefont {M.}~\bibnamefont
  {Fernandez-Gonzalez}}, \bibinfo {author} {\bibfnamefont {J.~A.}\ \bibnamefont
  {Gallegos}}, \bibinfo {author} {\bibfnamefont {E.~D.}\ \bibnamefont
  {Manuel}}, \bibinfo {author} {\bibfnamefont {F.~J.}\ \bibnamefont
  {Perales}},\ and\ \bibinfo {author} {\bibfnamefont {J.}~\bibnamefont
  {Leach}},\ }\bibfield  {title} {\bibinfo {title} {Images of science linked to
  labwork: A survey of secondary school and university students},\ }\href
  {https://link.springer.com/article/10.1023/A:1013141706723} {\bibfield
  {journal} {\bibinfo  {journal} {Research in Science Education}\ }\textbf
  {\bibinfo {volume} {31}},\ \bibinfo {pages} {499} (\bibinfo {year}
  {2001})}\BibitemShut {NoStop}%
\bibitem [{\citenamefont {Evangelinos}\ \emph {et~al.}(2002)\citenamefont
  {Evangelinos}, \citenamefont {Psillos},\ and\ \citenamefont
  {Valassiades}}]{Evangelinos2002}%
  \BibitemOpen
  \bibfield  {author} {\bibinfo {author} {\bibfnamefont {D.}~\bibnamefont
  {Evangelinos}}, \bibinfo {author} {\bibfnamefont {D.}~\bibnamefont
  {Psillos}},\ and\ \bibinfo {author} {\bibfnamefont {O.}~\bibnamefont
  {Valassiades}},\ }\bibinfo {title} {An investigation of teaching and learning
  about measurement data and their treatment in the introductory physics
  laboratory},\ in\ \href {https://doi.org/10.1007/0-306-48196-0_19} {\emph
  {\bibinfo {booktitle} {Teaching and Learning in the Science Laboratory}}},\
  \bibinfo {editor} {edited by\ \bibinfo {editor} {\bibfnamefont
  {D.}~\bibnamefont {Psillos}}\ and\ \bibinfo {editor} {\bibfnamefont
  {H.}~\bibnamefont {Niedderer}}}\ (\bibinfo  {publisher} {Springer
  Netherlands},\ \bibinfo {address} {Dordrecht},\ \bibinfo {year} {2002})\ pp.\
  \bibinfo {pages} {179--190}\BibitemShut {NoStop}%
\bibitem [{\citenamefont {Stein}\ \emph {et~al.}(2019)\citenamefont {Stein},
  \citenamefont {White}, \citenamefont {Passante},\ and\ \citenamefont
  {Holmes}}]{Stein2019}%
  \BibitemOpen
  \bibfield  {author} {\bibinfo {author} {\bibfnamefont {M.~M.}\ \bibnamefont
  {Stein}}, \bibinfo {author} {\bibfnamefont {C.}~\bibnamefont {White}},
  \bibinfo {author} {\bibfnamefont {G.}~\bibnamefont {Passante}},\ and\
  \bibinfo {author} {\bibfnamefont {N.~G.}\ \bibnamefont {Holmes}},\ }\bibfield
   {title} {\bibinfo {title} {Student interpretations of uncertainty in
  classical and quantum mechanics experiments},\ }in\ \href
  {https://www.compadre.org/per/items/detail.cfm?ID=15307} {\emph {\bibinfo
  {booktitle} {Physics Education Research Conference 2019}}},\ \bibinfo {series
  and number} {PER Conference}\ (\bibinfo {address} {Provo, UT},\ \bibinfo
  {year} {2019})\ pp.\ \bibinfo {pages} {573--578}\BibitemShut {NoStop}%
\bibitem [{\citenamefont {Stump}\ \emph {et~al.}(2020)\citenamefont {Stump},
  \citenamefont {White}, \citenamefont {Passante},\ and\ \citenamefont
  {Holmes}}]{Stump2020}%
  \BibitemOpen
  \bibfield  {author} {\bibinfo {author} {\bibfnamefont {E.~M.}\ \bibnamefont
  {Stump}}, \bibinfo {author} {\bibfnamefont {C.}~\bibnamefont {White}},
  \bibinfo {author} {\bibfnamefont {G.}~\bibnamefont {Passante}},\ and\
  \bibinfo {author} {\bibfnamefont {N.~G.}\ \bibnamefont {Holmes}},\ }\bibfield
   {title} {\bibinfo {title} {Student reasoning about sources of experimental
  measurement uncertainty in quantum versus classical mechanics},\ }in\
  \href@noop {} {\emph {\bibinfo {booktitle} {Physics Education Research
  Conference 2020}}},\ \bibinfo {series and number} {PER Conference}\ (\bibinfo
  {address} {Virtual Conference},\ \bibinfo {year} {2020})\ pp.\ \bibinfo
  {pages} {527--532}\BibitemShut {NoStop}%
\bibitem [{\citenamefont {White}\ \emph {et~al.}(2020)\citenamefont {White},
  \citenamefont {Stump}, \citenamefont {Holmes},\ and\ \citenamefont
  {Passante}}]{White2020}%
  \BibitemOpen
  \bibfield  {author} {\bibinfo {author} {\bibfnamefont {C.}~\bibnamefont
  {White}}, \bibinfo {author} {\bibfnamefont {E.~M.}\ \bibnamefont {Stump}},
  \bibinfo {author} {\bibfnamefont {N.~G.}\ \bibnamefont {Holmes}},\ and\
  \bibinfo {author} {\bibfnamefont {G.}~\bibnamefont {Passante}},\ }\bibfield
  {title} {\bibinfo {title} {Student evaluation of more or better experimental
  data in classical and quantum mechanics},\ }in\ \href@noop {} {\emph
  {\bibinfo {booktitle} {Physics Education Research Conference 2020}}},\
  \bibinfo {series and number} {PER Conference}\ (\bibinfo {address} {Virtual
  Conference},\ \bibinfo {year} {2020})\ pp.\ \bibinfo {pages}
  {575--580}\BibitemShut {NoStop}%
\bibitem [{\citenamefont {Stump}\ \emph {et~al.}(2023)\citenamefont {Stump},
  \citenamefont {Dew}, \citenamefont {Passante},\ and\ \citenamefont
  {Holmes}}]{StumpTBD}%
  \BibitemOpen
  \bibfield  {author} {\bibinfo {author} {\bibfnamefont {E.~M.}\ \bibnamefont
  {Stump}}, \bibinfo {author} {\bibfnamefont {M.}~\bibnamefont {Dew}}, \bibinfo
  {author} {\bibfnamefont {G.}~\bibnamefont {Passante}},\ and\ \bibinfo
  {author} {\bibfnamefont {N.~G.}\ \bibnamefont {Holmes}},\ }\href@noop {}
  {\bibinfo {title} {Context affects student thinking about sources of
  uncertainty in classical and quantum mechanics}} (\bibinfo {year} {2023}),\
  \Eprint {https://arxiv.org/abs/2306.14994} {arXiv:2306.14994 [physics.ed-ph]}
  \BibitemShut {NoStop}%
\bibitem [{\citenamefont {Schang}\ \emph {et~al.}(2023)\citenamefont {Schang},
  \citenamefont {Dew}, \citenamefont {Stump}, \citenamefont {Holmes},\ and\
  \citenamefont {Passante}}]{SchangTBD}%
  \BibitemOpen
  \bibfield  {author} {\bibinfo {author} {\bibfnamefont {A.}~\bibnamefont
  {Schang}}, \bibinfo {author} {\bibfnamefont {M.}~\bibnamefont {Dew}},
  \bibinfo {author} {\bibfnamefont {E.~M.}\ \bibnamefont {Stump}}, \bibinfo
  {author} {\bibfnamefont {N.~G.}\ \bibnamefont {Holmes}},\ and\ \bibinfo
  {author} {\bibfnamefont {G.}~\bibnamefont {Passante}},\ }\href@noop {}
  {\bibinfo {title} {New perspectives on student reasoning about measurement
  uncertainty: More or better data}} (\bibinfo {year} {2023}),\ \Eprint
  {https://arxiv.org/abs/2306.16975} {arXiv:2306.16975 [physics.ed-ph]}
  \BibitemShut {NoStop}%
\bibitem [{\citenamefont {Leach}\ \emph {et~al.}(2000)\citenamefont {Leach},
  \citenamefont {Millar}, \citenamefont {Ryder},\ and\ \citenamefont
  {S\'{e}r\'{e}}}]{Leach2000}%
  \BibitemOpen
  \bibfield  {author} {\bibinfo {author} {\bibfnamefont {J.}~\bibnamefont
  {Leach}}, \bibinfo {author} {\bibfnamefont {R.}~\bibnamefont {Millar}},
  \bibinfo {author} {\bibfnamefont {J.}~\bibnamefont {Ryder}},\ and\ \bibinfo
  {author} {\bibfnamefont {M.-G.}\ \bibnamefont {S\'{e}r\'{e}}},\ }\bibfield
  {title} {\bibinfo {title} {Epistemological understanding in science learning:
  the consistency of representations across contexts},\ }\href
  {https://doi.org/https://doi.org/10.1016/S0959-4752(00)00013-X} {\bibfield
  {journal} {\bibinfo  {journal} {Learning and Instruction}\ }\textbf {\bibinfo
  {volume} {10}},\ \bibinfo {pages} {497} (\bibinfo {year} {2000})}\BibitemShut
  {NoStop}%
\bibitem [{\citenamefont {Landis}\ and\ \citenamefont
  {Koch}(1977)}]{Landis1977}%
  \BibitemOpen
  \bibfield  {author} {\bibinfo {author} {\bibfnamefont {J.~R.}\ \bibnamefont
  {Landis}}\ and\ \bibinfo {author} {\bibfnamefont {G.~G.}\ \bibnamefont
  {Koch}},\ }\bibfield  {title} {\bibinfo {title} {The measurement of observer
  agreement for categorical data},\ }\href {https://doi.org/10.2307/2529310}
  {\bibfield  {journal} {\bibinfo  {journal} {Biometrics}\ }\textbf {\bibinfo
  {volume} {33}},\ \bibinfo {pages} {159} (\bibinfo {year} {1977})}\BibitemShut
  {NoStop}%
\bibitem [{\citenamefont {Zwickl}\ \emph {et~al.}(2015)\citenamefont {Zwickl},
  \citenamefont {Hu}, \citenamefont {Finkelstein},\ and\ \citenamefont
  {Lewandowski}}]{Zwickl2015}%
  \BibitemOpen
  \bibfield  {author} {\bibinfo {author} {\bibfnamefont {B.~M.}\ \bibnamefont
  {Zwickl}}, \bibinfo {author} {\bibfnamefont {D.}~\bibnamefont {Hu}}, \bibinfo
  {author} {\bibfnamefont {N.}~\bibnamefont {Finkelstein}},\ and\ \bibinfo
  {author} {\bibfnamefont {H.~J.}\ \bibnamefont {Lewandowski}},\ }\bibfield
  {title} {\bibinfo {title} {Model-based reasoning in the physics laboratory:
  Framework and initial results},\ }\href
  {https://link.aps.org/doi/10.1103/PhysRevSTPER.11.020113} {\bibfield
  {journal} {\bibinfo  {journal} {Phys. Rev. ST Phys. Educ. Res.}\ }\textbf
  {\bibinfo {volume} {11}},\ \bibinfo {pages} {020113} (\bibinfo {year}
  {2015})}\BibitemShut {NoStop}%
\bibitem [{\citenamefont {Dounas-Frazer}\ and\ \citenamefont
  {Lewandowski}(2018)}]{DounasFrazer2018}%
  \BibitemOpen
  \bibfield  {author} {\bibinfo {author} {\bibfnamefont {D.~R.}\ \bibnamefont
  {Dounas-Frazer}}\ and\ \bibinfo {author} {\bibfnamefont {H.~J.}\ \bibnamefont
  {Lewandowski}},\ }\bibfield  {title} {\bibinfo {title} {The modelling
  framework for experimental physics: description, development, and
  applications},\ }\href@noop {} {\bibfield  {journal} {\bibinfo  {journal}
  {European Journal of Physics}\ }\textbf {\bibinfo {volume} {39}},\ \bibinfo
  {pages} {064005} (\bibinfo {year} {2018})}\BibitemShut {NoStop}%
\bibitem [{\citenamefont {Wilson}(1927)}]{Wilson1927}%
  \BibitemOpen
  \bibfield  {author} {\bibinfo {author} {\bibfnamefont {E.~B.}\ \bibnamefont
  {Wilson}},\ }\bibfield  {title} {\bibinfo {title} {Probable inference, the
  law of succession, and statistical inference},\ }\href
  {https://doi.org/10.1080/01621459.1927.10502953} {\bibfield  {journal}
  {\bibinfo  {journal} {Journal of the American Statistical Association}\
  }\textbf {\bibinfo {volume} {22}},\ \bibinfo {pages} {209} (\bibinfo {year}
  {1927})}\BibitemShut {NoStop}%
\bibitem [{\citenamefont {Cohen}(1994)}]{cohen_earth_1994}%
  \BibitemOpen
  \bibfield  {author} {\bibinfo {author} {\bibfnamefont {J.}~\bibnamefont
  {Cohen}},\ }\bibfield  {title} {\bibinfo {title} {The earth is round (p
  {\textless} .05).},\ }\href {https://doi.org/10.1037/0003-066X.49.12.997}
  {\bibfield  {journal} {\bibinfo  {journal} {American Psychologist}\ }\textbf
  {\bibinfo {volume} {49}},\ \bibinfo {pages} {997} (\bibinfo {year}
  {1994})}\BibitemShut {NoStop}%
\bibitem [{\citenamefont {Cumming}(2013)}]{cumming_new_2013}%
  \BibitemOpen
  \bibfield  {author} {\bibinfo {author} {\bibfnamefont {G.}~\bibnamefont
  {Cumming}},\ }\bibfield  {title} {\bibinfo {title} {The {New} {Statistics}:
  {Why} and {How}},\ }\href {https://doi.org/10.1177/0956797613504966}
  {\bibfield  {journal} {\bibinfo  {journal} {Psychological Science}\ }\textbf
  {\bibinfo {volume} {25}},\ \bibinfo {pages} {7} (\bibinfo {year}
  {2013})}\BibitemShut {NoStop}%
\bibitem [{\citenamefont {Nosek}\ \emph {et~al.}(2018)\citenamefont {Nosek},
  \citenamefont {Ebersole}, \citenamefont {DeHaven},\ and\ \citenamefont
  {Mellor}}]{nosek_preregistration_2018}%
  \BibitemOpen
  \bibfield  {author} {\bibinfo {author} {\bibfnamefont {B.~A.}\ \bibnamefont
  {Nosek}}, \bibinfo {author} {\bibfnamefont {C.~R.}\ \bibnamefont {Ebersole}},
  \bibinfo {author} {\bibfnamefont {A.~C.}\ \bibnamefont {DeHaven}},\ and\
  \bibinfo {author} {\bibfnamefont {D.~T.}\ \bibnamefont {Mellor}},\ }\bibfield
   {title} {\bibinfo {title} {The preregistration revolution.},\ }\href
  {https://doi.org/10.1073/pnas.1708274114} {\bibfield  {journal} {\bibinfo
  {journal} {Proceedings of the National Academy of Sciences of the United
  States of America}\ }\textbf {\bibinfo {volume} {115}},\ \bibinfo {pages}
  {2600} (\bibinfo {year} {2018})},\ \bibinfo {note} {publisher: National
  Academy of Sciences}\BibitemShut {NoStop}%
\bibitem [{\citenamefont {Wilson}\ \emph {et~al.}(2022)\citenamefont {Wilson},
  \citenamefont {Pollard}, \citenamefont {Aiken}, \citenamefont {Caballero},\
  and\ \citenamefont {Lewandowski}}]{Wilson2022}%
  \BibitemOpen
  \bibfield  {author} {\bibinfo {author} {\bibfnamefont {J.}~\bibnamefont
  {Wilson}}, \bibinfo {author} {\bibfnamefont {B.}~\bibnamefont {Pollard}},
  \bibinfo {author} {\bibfnamefont {J.~M.}\ \bibnamefont {Aiken}}, \bibinfo
  {author} {\bibfnamefont {M.~D.}\ \bibnamefont {Caballero}},\ and\ \bibinfo
  {author} {\bibfnamefont {H.~J.}\ \bibnamefont {Lewandowski}},\ }\bibfield
  {title} {\bibinfo {title} {Classification of open-ended responses to a
  research-based assessment using natural language processing},\ }\href
  {https://doi.org/10.1103/PhysRevPhysEducRes.18.010141} {\bibfield  {journal}
  {\bibinfo  {journal} {Phys. Rev. Phys. Educ. Res.}\ }\textbf {\bibinfo
  {volume} {18}},\ \bibinfo {pages} {010141} (\bibinfo {year}
  {2022})}\BibitemShut {NoStop}%
\end{thebibliography}%

\appendix

\section{Survey questions}

\begin{figure*}
    \centering
    \includegraphics[width=.8\textwidth]{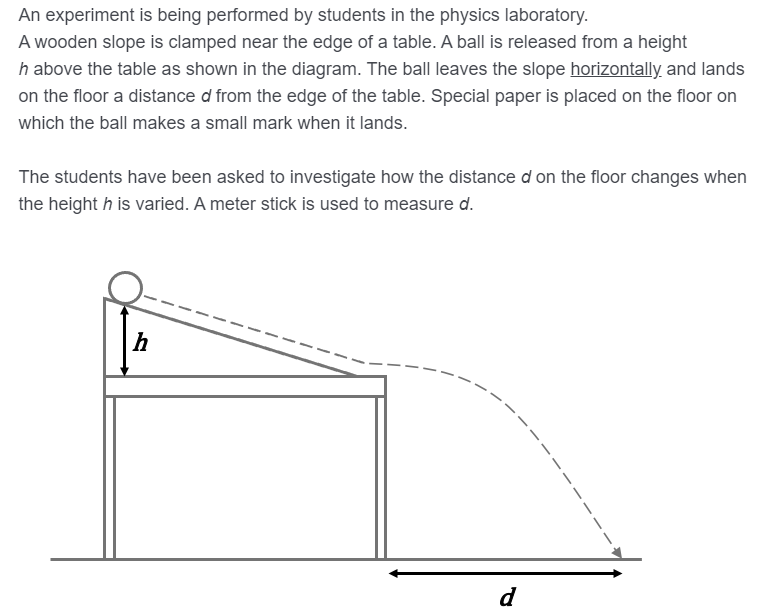}
    \caption{The overall experimental scenario, modified from ~\cite{Allie1998}.}
    \label{fig:scenario}
\end{figure*}

\begin{figure*}
    \centering
    \includegraphics[width=.8\textwidth]{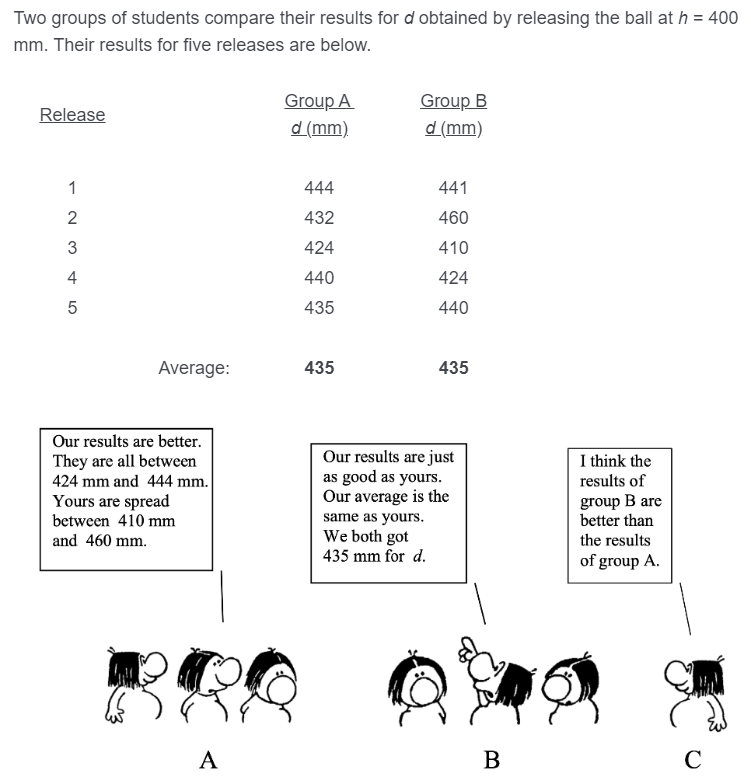}
    \caption{The SMDS probe, modified from ~\cite{Allie1998}.}
    \label{fig:smds}
\end{figure*}

\begin{figure*}
    \centering
    \includegraphics[width=.8\textwidth]{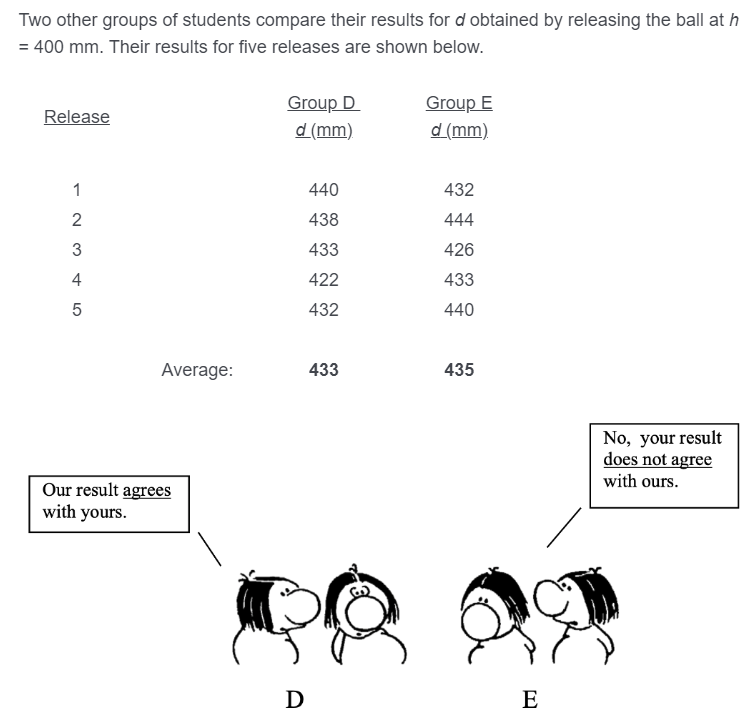}
    \caption{The DMSS probe, modified from ~\cite{Allie1998}.}
    \label{fig:dmss}
\end{figure*}

A visual representation of the SMDS and DMSS survey questions is shown in Figs.~\ref{fig:scenario},~\ref{fig:smds}, and~\ref{fig:dmss}.

\section{Demographic information}

The self-reported demographic information for the survey participants is shown in Table~\ref{ta:demographics}.
\begin{table}[tb] 
  \caption{Demographic information self-reported by the students included in this study (427 intro students and 158 beyond-intro students). Students who marked two or more races are counted in each race category they chose.}
  \label{ta:demographics}
  \begin{ruledtabular}
    \begin{tabular}{lrr}
    & Intro & Beyond-intro\\
    \hline
    \textbf{Year of college} \\
    \quad First year (freshman) & 191 & 1\\
    \quad Second year (sophomore) & 156 & 28\\
    \quad Third year (junior) & 42 & 63\\
    \quad Fourth year + (senior) & 23 & 56\\
    \quad Graduate student & 0 & 2\\
    \quad Unspecified & 15 & 8 \\
    \textbf{Gender}\\
    \quad Female & 163 & 37\\
    \quad Male & 241 & 110\\
    \quad Non-binary & 2 & 4\\
    \quad Unspecified & 21 & 7\\
    \textbf{Race/ethnicity}\\
    \quad American Indian or Alaska Native & 6 & 3\\
    \quad Asian or Asian American & 78 & 34 \\
    \quad Black or African American & 83 & 4 \\
    \quad Hispanic or Latinx & 58 & 21 \\
    \quad Native Hawaiian or other Pacific Islander & 4 & 2\\
    \quad Prefer to self-describe & 3 & 4\\
    \quad White & 214 & 105\\
    \quad Unspecified & 18 & 9\\
    \textbf{First-generation status}\\
    \quad First-generation college student & 77 & 26\\
    \quad Not first-generation college student & 327 & 125\\
    \quad Unspecified & 23 & 7\\
    \end{tabular}
  \end{ruledtabular}
\end{table}

\section{Additional comparisons across student survey responses}
\label{sec:figs}
\begin{figure}
  \centering
  \subfloat[]{\includegraphics[width=.48\textwidth]{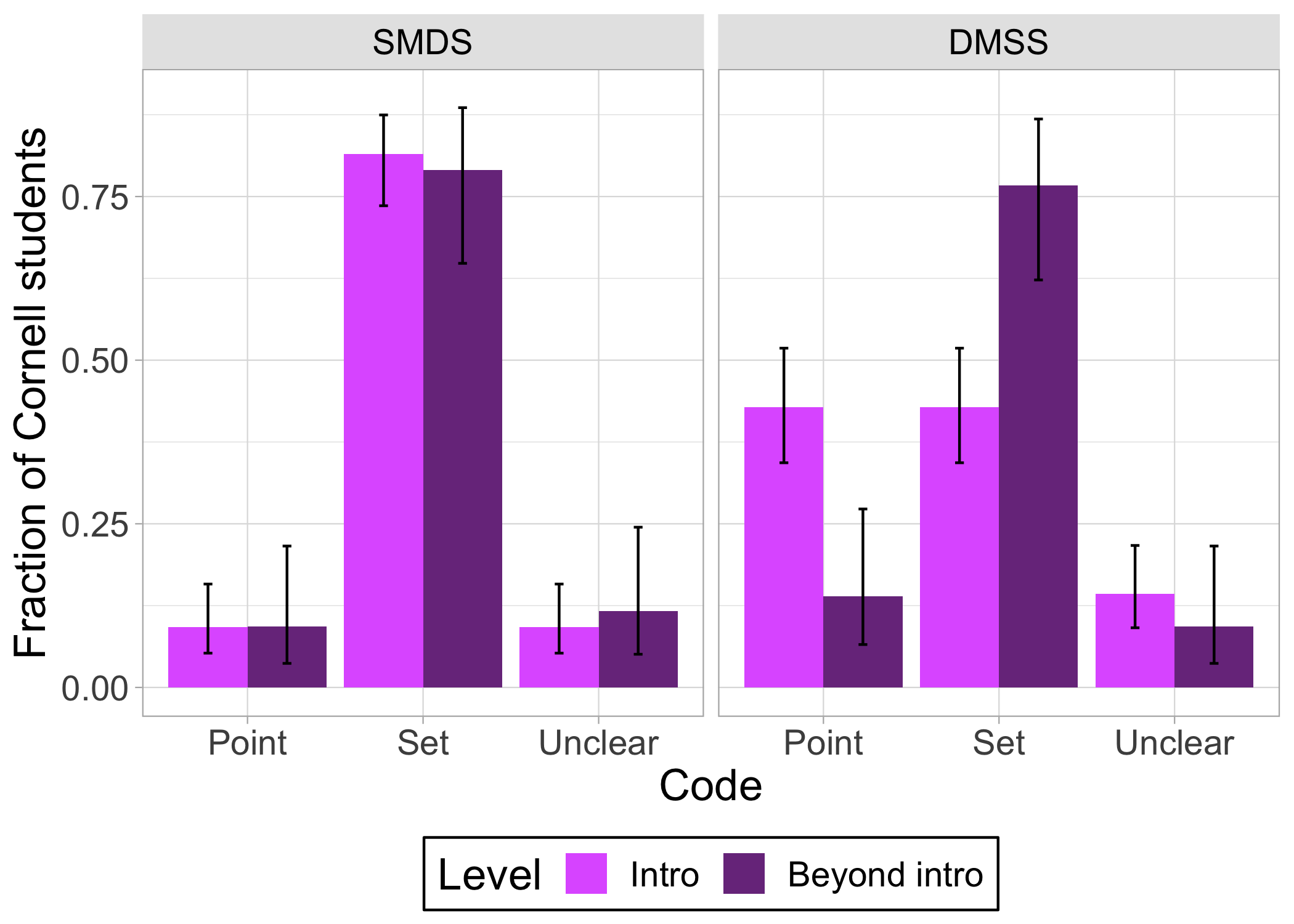}}\\
  \subfloat[]{\includegraphics[width=.48\textwidth]{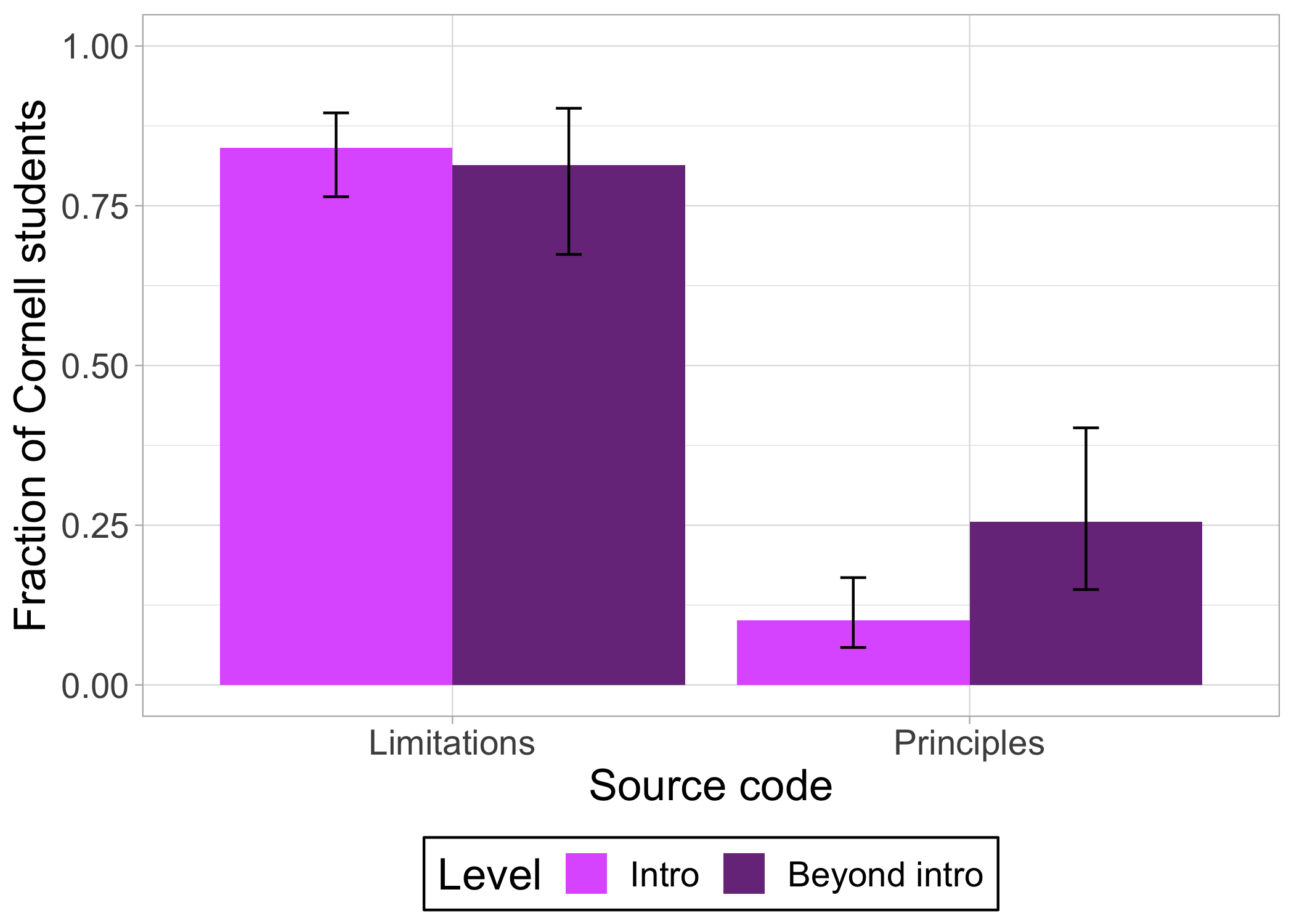}}\\
  \subfloat[]{\includegraphics[width=.48\textwidth]{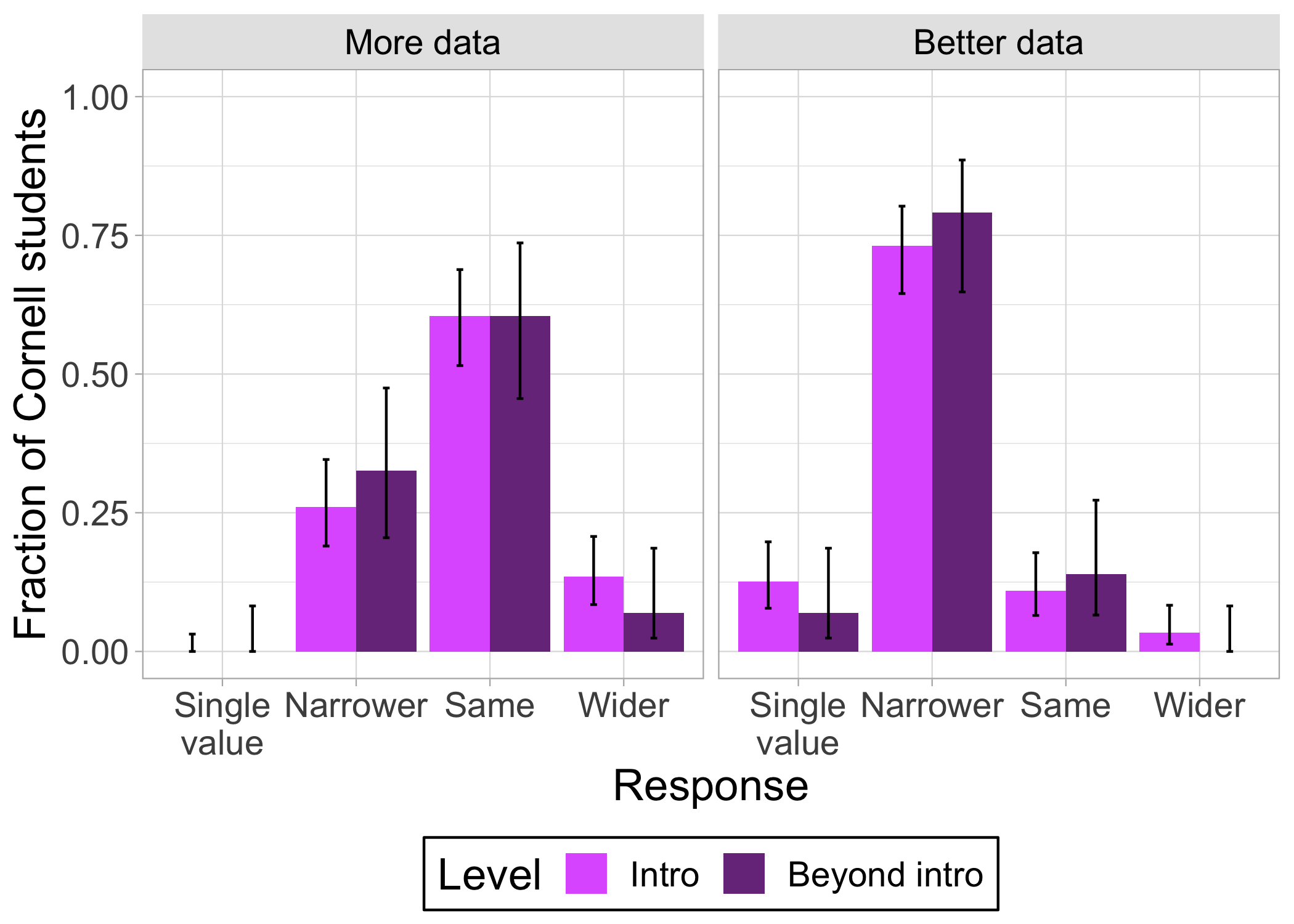}}\\
  \caption{Comparison of intro and beyond-intro Cornell students' responses to the PMQ probes (top), sources of uncertainty question (middle), and more/better data questions (bottom). Uncertainty bars represent the 95\% confidence interval.}
    \label{fig:Cornell}
\end{figure}

First, we wanted to confirm that the differences between intro and beyond-intro students were not exclusively explained by institution differences. To do so, we compared intro and beyond-intro students' responses within a single university, Cornell. These comparisons are shown in Fig.~\ref{fig:Cornell}.
We observed that most of the trends present in the full data set were reflected in the Cornell results. Within the Cornell population, we saw no differences in intro and beyond-intro students' responses to the SMDS probe, listing of limitations sources of uncertainty, and responses to the More Data question, as in the full data set. We saw differences in student's responses to the DMSS probe and listing of principles sources of uncertainty similar to the full data set. The only discrepancy in conclusions we would draw in the Cornell-only data set compared to the full data set is for the Better Data question. In the full data, we observed a difference between intro and beyond-intro students' responses, but within the Cornell data these two groups' responses are indistinguishable. However, the trends in the observed fractions for each answer response align with the full-data results, even if the fractions are indistinguishable within uncertainty: a larger fraction of intro students than beyond-intro students answered Single value (13\% and 7\%, respectively), while a smaller fraction of intro students than beyond-intro students answered Narrower (73\% and 79\%, respectively). Overall, therefore, the Cornell-specific results agree with the full-data results.

\begin{figure}
    \centering
    \subfloat[]{\includegraphics[width=.48\textwidth]{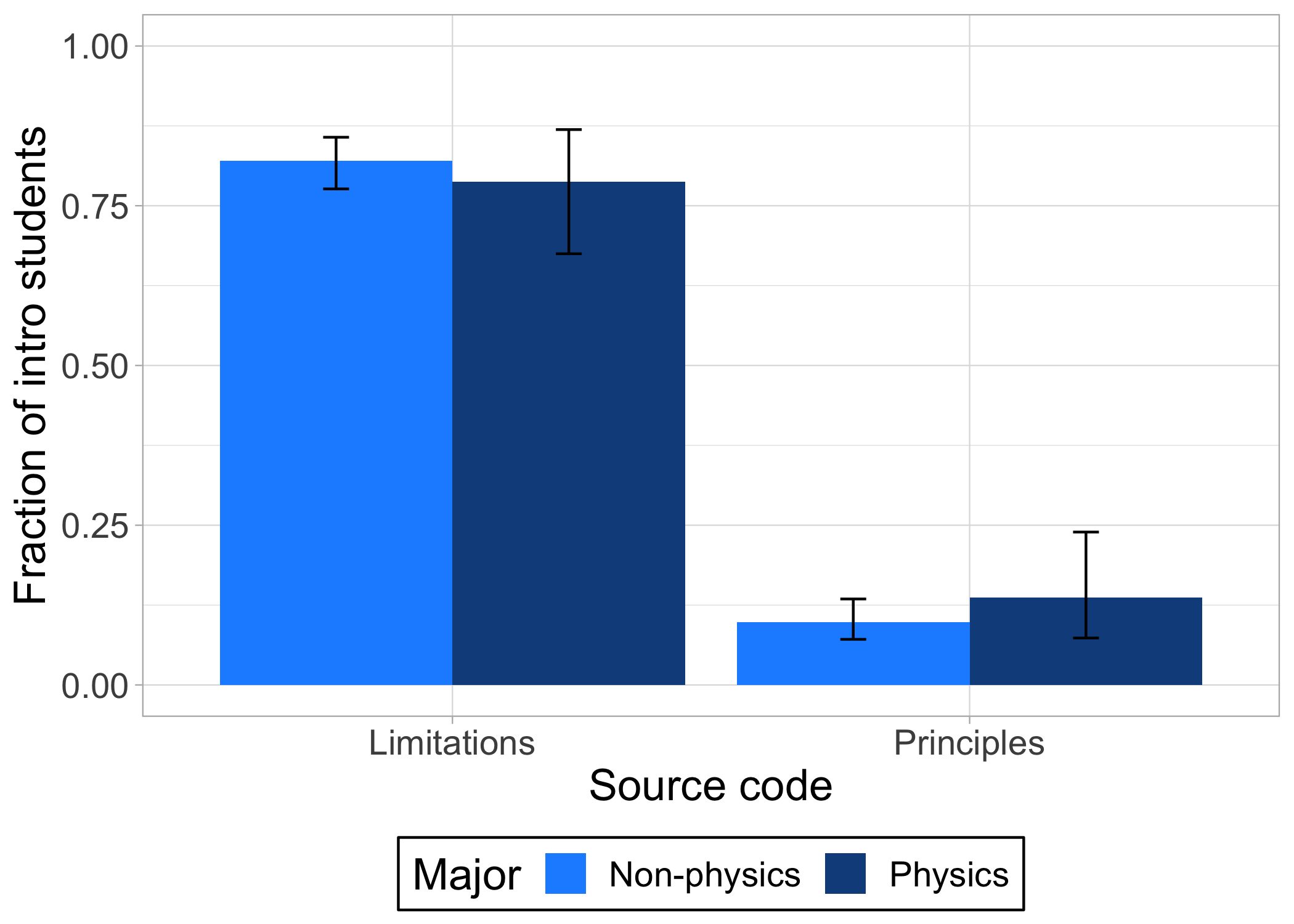}}\\
    \subfloat[]{\includegraphics[width=.48\textwidth]{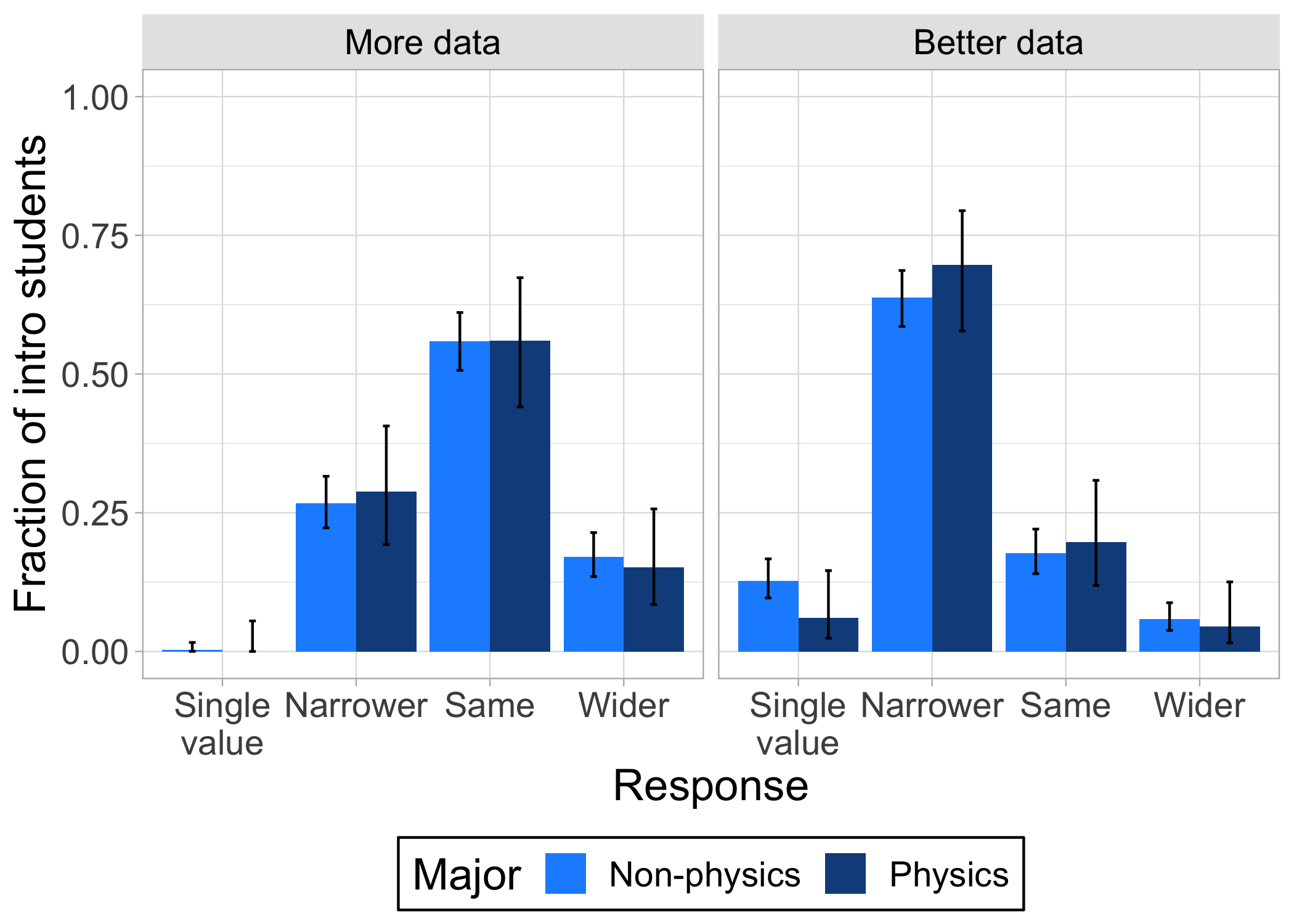}}
    \caption{Comparison of intro physics majors' and intro non-physics majors' responses to the sources of uncertainty question (top) and more/better data questions (bottom). Uncertainty bars represent the 95\% confidence interval.}
    \label{fig:major}
\end{figure}

To test whether differences between intro and beyond-intro students' responses were due to differences in student major, we compared intro physics majors' and intro non-physics majors' responses. These comparisons for the Sources, More Data, and Better Data questions are shown in Fig.~\ref{fig:major}. For these three questions, we observed no differences in intro students' responses based on major.

\begin{figure}
    \centering
    \subfloat[]{\includegraphics[width=.48\textwidth]{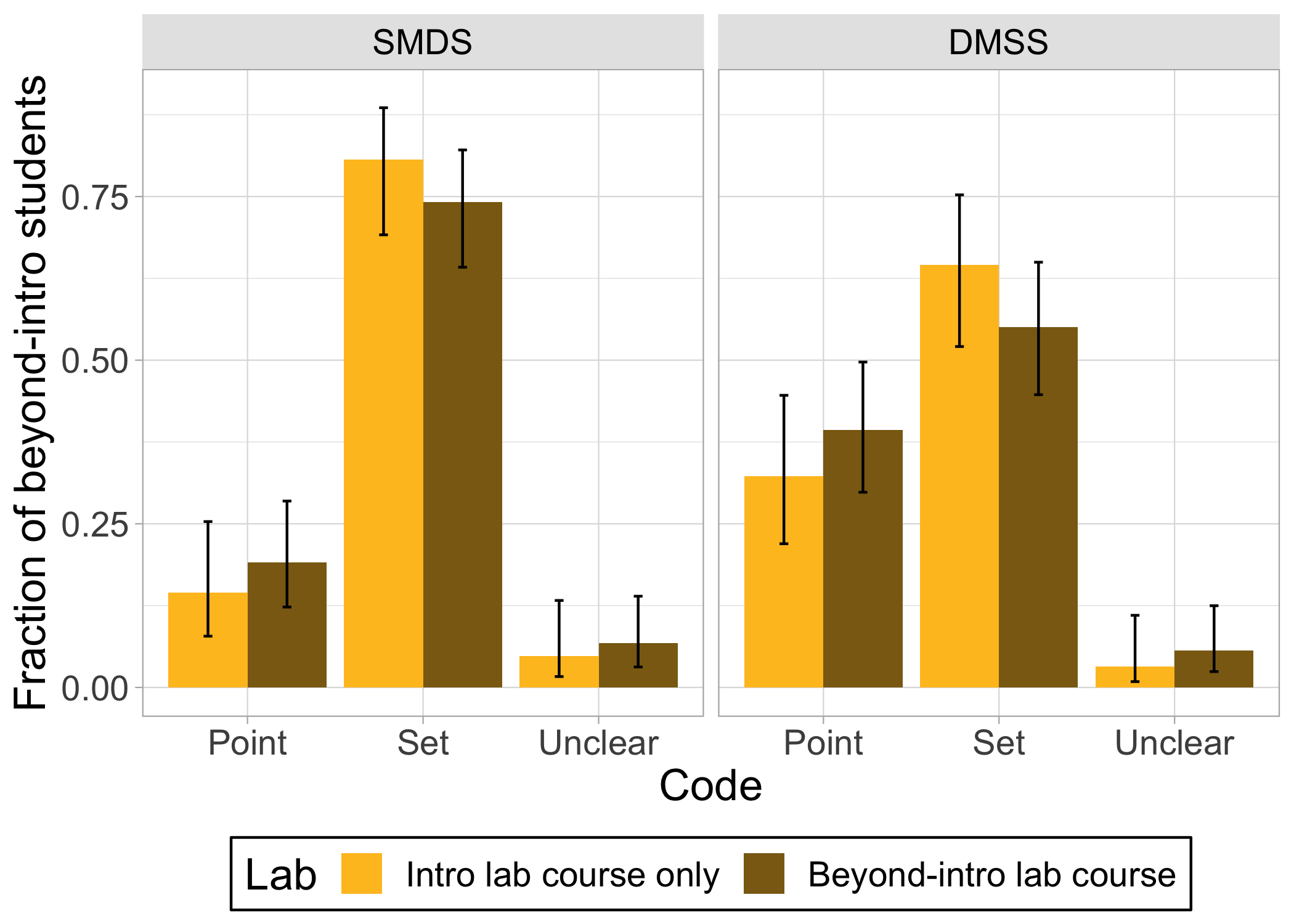}}\\
    \subfloat[]{\includegraphics[width=.48\textwidth]{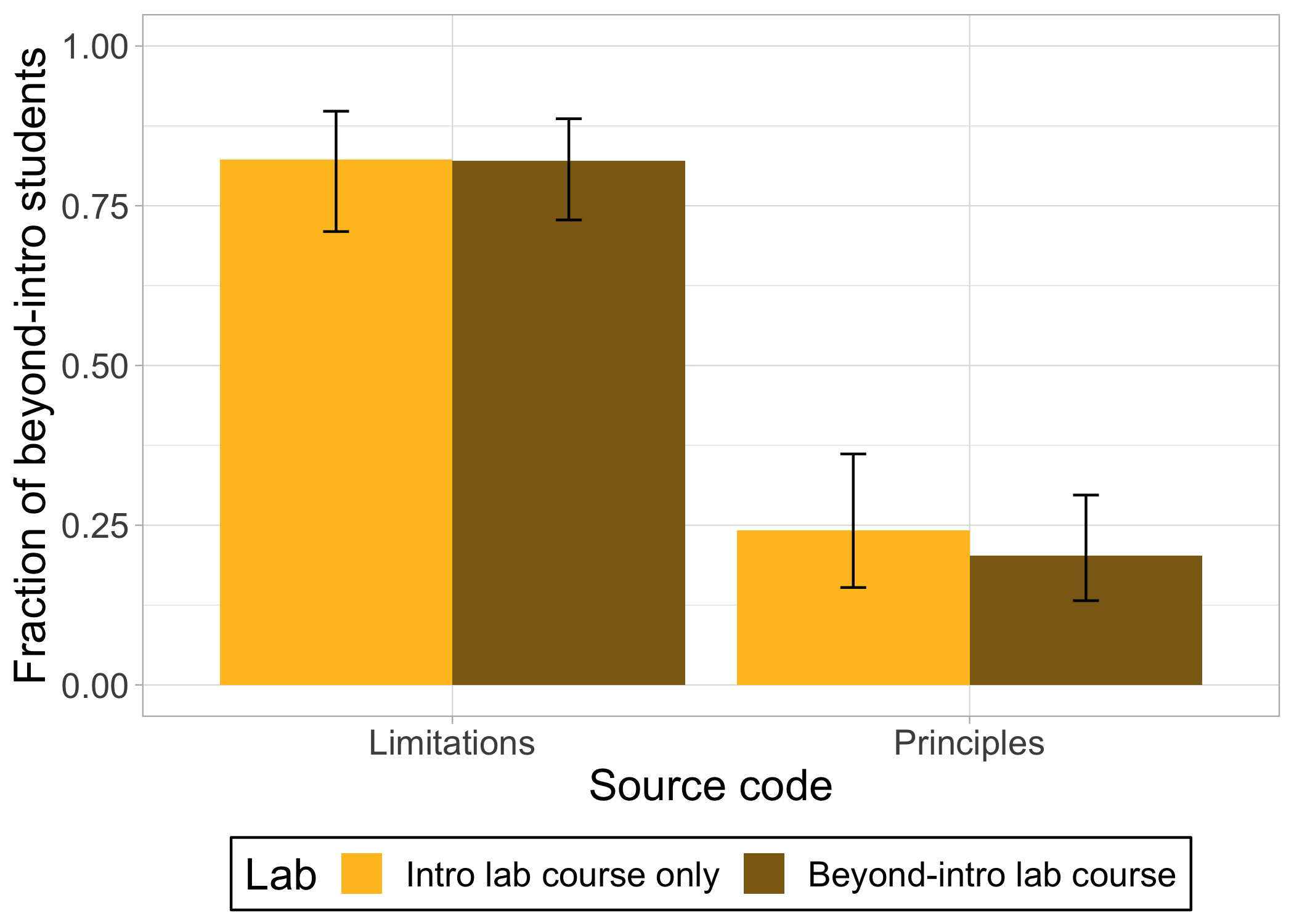}}\\
    \subfloat[]{\includegraphics[width=.48\textwidth]{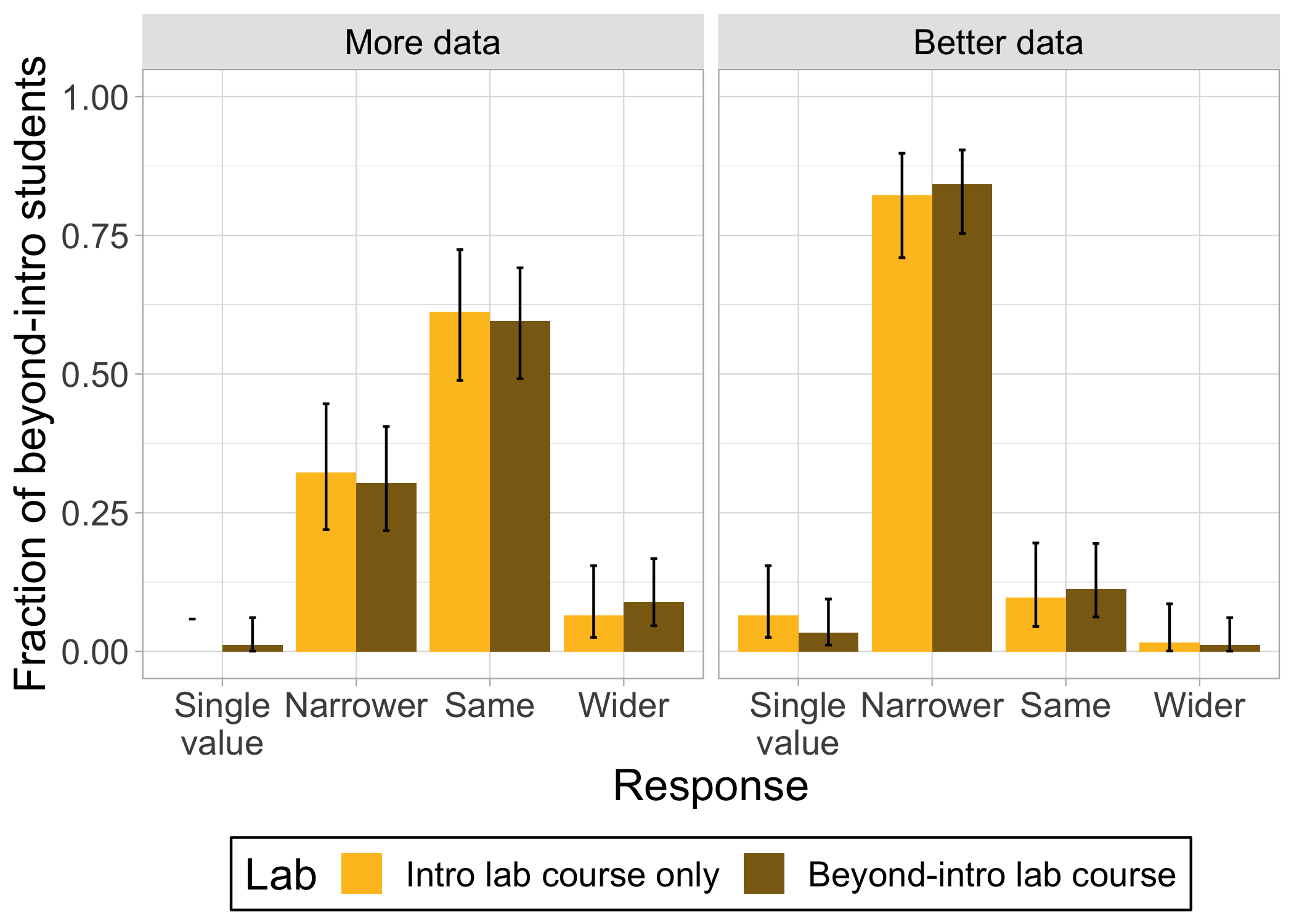}}
    \caption{Comparison of beyond-intro students' responses to the PMQ probes (top), sources of uncertainty question (middle), and more/better data questions (bottom) based on what lab courses students had taken. Uncertainty bars represent the 95\% confidence interval.}
    \label{fig:lab}
\end{figure}

To test whether differences between intro and beyond-intro students' responses were due to differences in what lab courses students had taken, we compared beyond-intro students' responses based on whether they had taken only intro-level lab courses or had taken (or were currently taking) at least one beyond-intro lab course. These comparisons are shown in Fig.~\ref{fig:lab}. We observed on differences in beyond-intro students' responses based on lab courses taken.

\begin{figure}
    \centering
    \subfloat[]{\includegraphics[width=.48\textwidth]{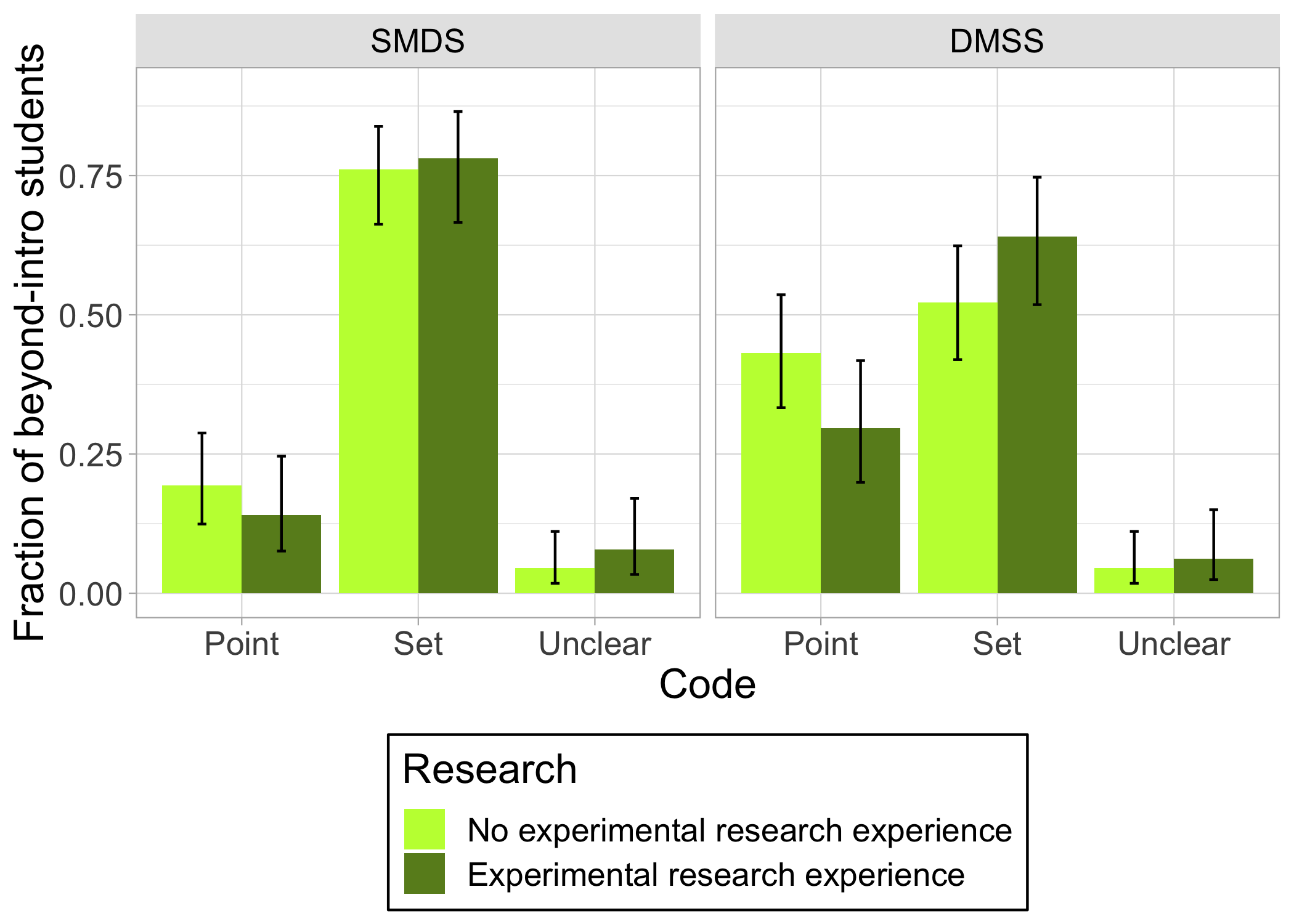}}\\
    \subfloat[]{\includegraphics[width=.48\textwidth]{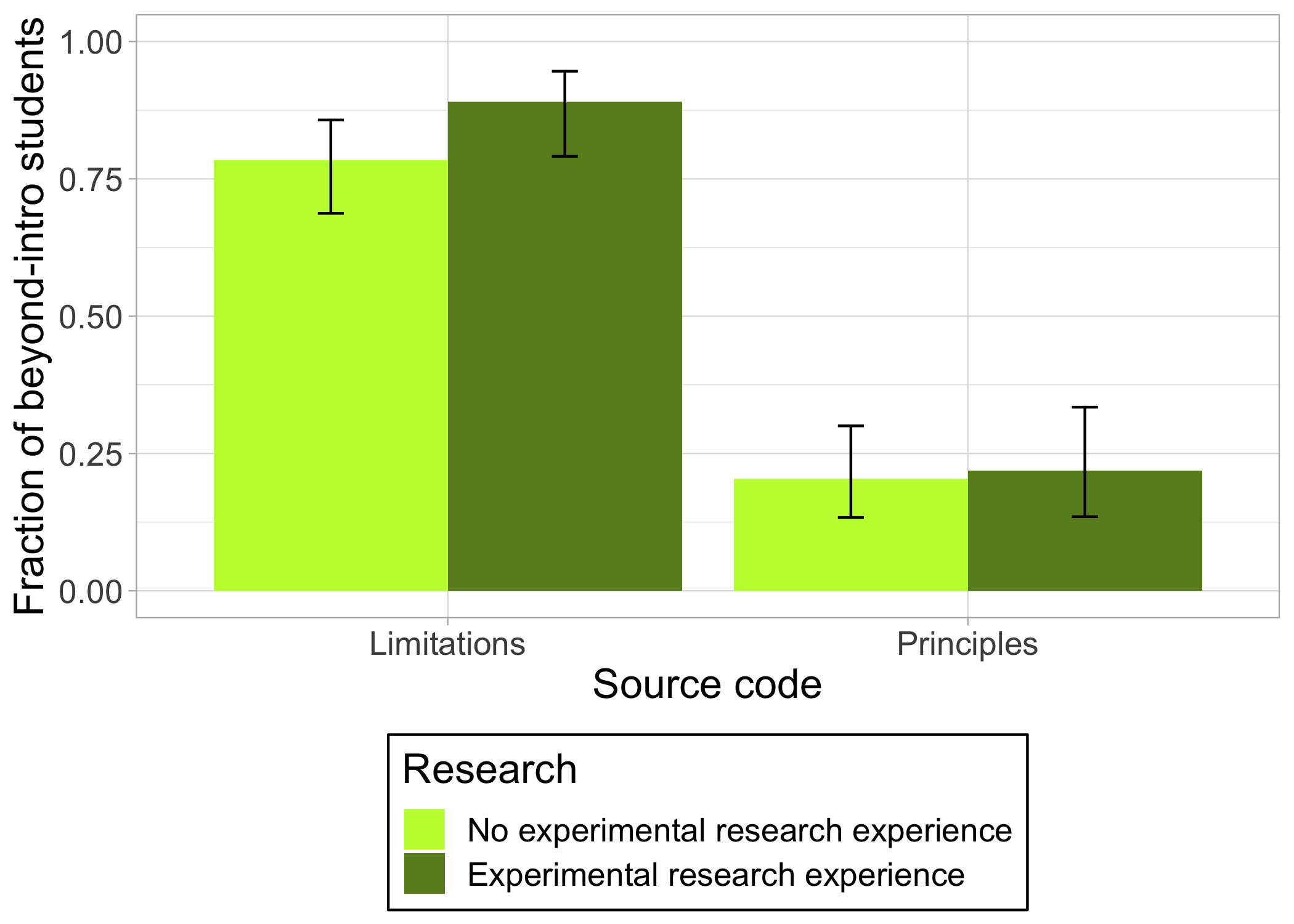}}\\
    \subfloat[]{\includegraphics[width=.48\textwidth]{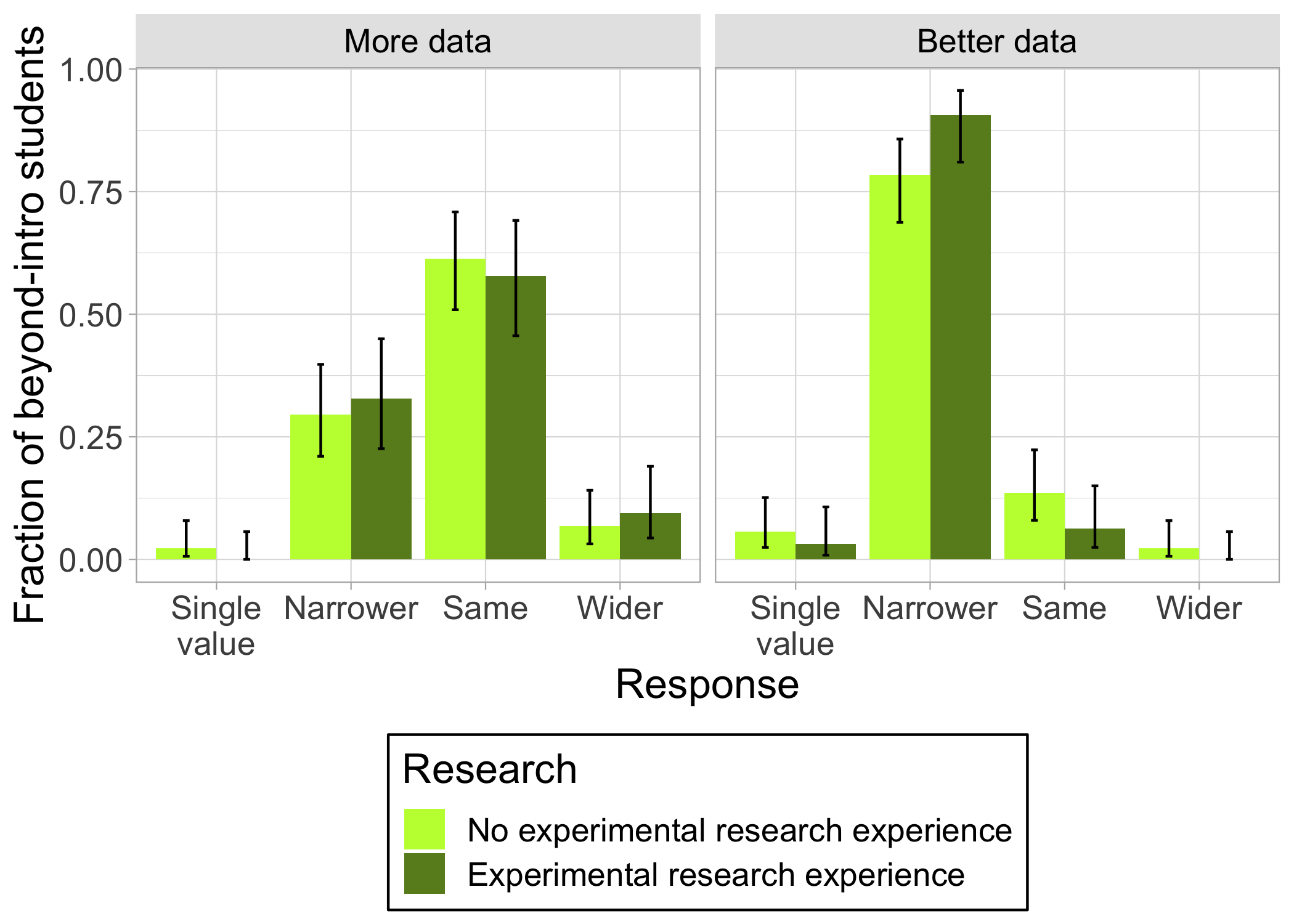}}
    \caption{Comparison of beyond-intro students' responses to the PMQ probes (top), sources of uncertainty question (middle), and more/better data questions (bottom) based on research experience. Uncertainty bars represent the 95\% confidence interval.}
    \label{fig:research}
\end{figure}

To test whether differences between intro and beyond-intro students' responses were due to differences in students' experience conducting research in an experimental lab setting, we compared beyond-intro students' responses based on whether they had experimental research experience. These comparisons are shown in Fig.~\ref{fig:research}. We observed on differences in beyond-intro students' responses based on research experience.

\end{document}